\documentclass[twocolumn,tight]{aastex631}
\usepackage[T1]{fontenc}

\def\mJB{{\rm mJy~beam^{-1}}}
\def\uJB{\mu {\rm Jy~beam^{-1}}}
\def\mJ{{\rm mJy}}

\def\Lsun{L_{\sun}}

\def\II{{\rm I\hspace{-0.1em}I}}

\shorttitle{Grain size in TMC-1A}
\shortauthors{Aso et al.}
\graphicspath{{./}{figures/}}

\begin{document}

\title{Grain Size in the Class I Protostellar System TMC-1A Constrained with ALMA and VLA Observations}

\author[0000-0002-8238-7709]{Yusuke Aso}
\affil{Korea Astronomy and Space Science Institute (KASI), 776 Daedeokdae-ro, Yuseong-gu, Daejeon 34055, Republic of Korea}

\author[0000-0002-9661-7958]{Satoshi Ohashi}
\affil{National Astronomical Observatory of Japan, 2-21-1 Osawa, Mitaka, Tokyo 181-8588, Japan}

\author[0000-0003-2300-2626]{Hauyu Baobab Liu}
\affil{Physics Department, National Sun Yat-Sen University, Taiwan, Republic of Chinese}
\affil{Center of Astronomy and Gravitation, National Taiwan Normal University, Taipei 116, Taiwan}

\author[0000-0002-9408-2857]{Wenrui Xu}
\affil{Center for Computational Astrophysics, Flatiron Institute, NY, USA}



\begin{abstract}
The disk mass and substructure in young stellar objects suggest that planet formation may start at the protostellar stage through the growth of dust grains. To accurately estimate the grain size at the protostellar stage, we have observed the Class I protostar TMC-1A using the Jansky Very Large Array (VLA) at the Q (7 mm) and Ka (9 mm) bands at a resolution of $\sim 0\farcs 2$ and analyzed archival data of Atacama Large Millimeter/submillimeter Array (ALMA) at Band 6 (1.3 mm) and 7 (0.9 mm) that cover the same spatial scale. The VLA images show a compact structure with a size of $\sim 25$~au and a spectral index of $\sim 2.5$. The ALMA images show compact and extended structures with a spectral index of $\sim 2$ at the central $\sim 40$~au region and another index of $\sim 3.3$ in the outer region. Our SED analysis using the observed fluxes at the four bands suggests one branch with a small grain size of $\sim 0.12$~mm and another with a grown grain size of $\sim 4$~mm. We also model polarized dust continuum emission adopting the two grain sizes and compare them with an observational result of TMC-1A, suggesting that the small grain size is preferable to the grown grain size. The small grain size implies gravitational instability in the TMC-1A disk, which is consistent with a spiral-like component recently identified.
\end{abstract}

\keywords{Circumstellar disks (235) --- Circumstellar grains (239) --- Protostars (1302) --- Low mass stars (2050)}

\section{Introduction} \label{sec:intro}

The earliest stage of planet formation in protoplanetary disks involves the growth of dust grains \citep[e.g.,][]{2014prpl.conf..339T}. The maximum grain size ($a_{\rm max}$) in these disks is a fundamental parameter in planet formation, as mechanisms like streaming instability \citep[e.g.,][]{2014prpl.conf..547J} and pebble accretion \citep[e.g.,][]{2020RAA....20..164L} are highly dependent on it \citep[e.g.,][]{2023ASPC..534..717D}.

During the formation of gas giant planets, these planets interact with the host disk, generating various substructures through planet-disk interactions \citep[e.g.,][]{2023ASPC..534..685P}. These substructures, including rings, gaps, spirals, and crescents, have been both predicted \citep[e.g.,][]{2023ASPC..534..423B} and observed in numerous disks around Class $\II$ objects, with rings and gaps being the most common \citep[e.g.,][]{2018ApJ...869L..41A}. While many of the planets inferred from disk gaps remain unconfirmed, the successful detection of the planets PDS 70 b and c within the gap of the PDS 70 disk \citep[][]{2018A&A...617A..44K, 2019NatAs...3..749H} strongly supports the link between these substructures and the presence of giant planets.

Similar to Class $\II$ disks, substructures have also been observed in some class 0/I disks \citep{2017ApJ...840L..12S, Sheehan2018ApJ...857...18S, SeguraCox2020Natur.586..228S, 2020NatAs...4..142L, Nakatani2020ApJ...895L...2N, 2023ApJ...956....9H, 2020ApJ...898...36L, 2024A&A...689L...5M}. These discoveries suggest that planet formation could begin as early as the protostellar stage. Examining dust grain growth at this early phase thus provides crucial insights into the initial conditions for planet formation.

Grain size measurement is necessary to verify whether dust grains can grow enough to start planet formation. Diagnosing long-wavelength excess in the spectral energy distribution (SED) has  been regarded a useful assessment of grain growth in disks and envelopes around young stellar objects \citep[e.g.,][]{1983QJRAS..24..267H, 1993Icar..106...20M, 2001ApJ...553..321D, 2006ApJ...636.1114D}, since a dust grain cannot efficiently emit/absorb radiation at wavelengths that are much longer than its diameter. Recent theoretical studies suggest an independent method through polarization due to self-scattering, which is applied to observational results to constrain the grain size in circumstellar disks \citep[e.g.,][]{2016ApJ...820...54K, 2017ApJ...844L...5K, 2018ApJ...864...81O, 2018ApJ...865L..12B, 2019ApJ...886..103O, 2019MNRAS.482L..29D, 2019ApJ...883...16M, 2020MNRAS.496..169L, 2021ApJ...910...75L, 2021ApJ...920...71A, 2022MNRAS.512.3922L, 2023ApJ...947L...5T}. Multi-band polarization observaiton can constrain the maximum grain size because, in the case of self-scattering, polarization fraction is highetst when the maximum grain size is around the observed wavelength over $2\pi$. Overall, longer, centimeter-wavelength observations are preferable for determining whether dust grains have grown to millimeter or centimeter sizes.

Grain size estimation, using the SED or self-scattering method, has been performed only with a limited number of protostellar disks. \citet{2024A&A...682A..56Z} reported presence of 1-mm-sized grains inside the soot line (within this line, temperature is high enough to sublimate solid amorphous refractory carbon) of the disk in the Class 0 protostellar system IRAS 16293-2422 B using ALMA (1.3 and 3 mm) and VLA (9 and 18 mm) continuum observations at angular resolutions of $\lesssim 0\farcs 2$. The self-scattering method has recently started to be applied to protostellar disks, in addition to the Class I/$\II$ disk in HL Tau and more evolved T Tauri disks: the grain size is estimated to be 50-$75~\micron$ in the Class 0 protostar HH212 \citep{2021ApJ...910...75L} and 80-$300~\micron$ in the Class I protostar TMC-1A \citep{2021ApJ...920...71A}. \citet{2019ApJS..245....2S} detected polarized continuum emission due to the self-scattering in nine protostellar systems in the Ophiuchus molecular cloud at the 1.3 mm wavelength in ALMA observations at an angular resolution of $\sim 0\farcs 25$.

TMC-1A is the Class I protostar with typical structures: disk, envelope, and outflow \citep[e.g., ][]{2015ApJ...812...27A, Bjerkeli2016Natur.540..406B} in the Turus star forming region at the distance of 141.8 pc \citep{2021AJ....162..110K}. A previous observational study suggested presence of mm-sized grains in its disk based on line absorption against strong continuum emission at the 1.3 mm wavelength \citep{Harsono2018NatAs...2..646H}. The authors discussed that the high opacity is preferred to high mass of the disk to explain the high optical depth because the TMC-1A disk does not exhibit a substructure as clearly as reported in Class $\II$ disks. Meanwhile, previous polarization observations at the 1.3 mm wavelength detected a component due to self-scattering, suggesting a grain size of $\sim 0.1$~mm \citep{2021ApJ...920...71A}. Multi-band observation is crucial to accurately estimate the grain size in this disk in the protostellar phase. For this purpose, we have observed TMC-1A using Karl G. Jansky Very Large Array (VLA) in the Q (7 mm) and Ka (9 mm) bands.
We also searched for ALMA archival data sets with similar $uv$-coverage to that of our VLA observations; as a result, ALMA Band 6 and 7 data sets are selected. The visibilities and images in the four bands are analyzed under coherent conditions.

The rest of the paper has the  following structure. Section~\ref{sec:obs} describes the details of our VLA observations and the selected ALMA archival data. Section~\ref{sec:result} shows the consistency of the visibilities and images at each band as well as spectral index maps. We analyze the spectral envergy distribution at seven positions, showing two solution branches in Section~\ref{sec:sed}. Further discussion about preference of the branches are in Section~\ref{sec:discussion}. A summary of our results and interpretation is in Section~\ref{sec:conclusion}. Appendix \ref{appendix:almadata} shows the detailes of ALMA data calibration.

\section{Observations and Data} \label{sec:obs}
\subsection{VLA Observations}
We observed TMC-1A using VLA with the antenna configuration of B in the Q and Ka bands (Project ID: 21B-089). The mapping center was $\alpha ({\rm J2000})=04^{\rm h}39^{\rm m}35\fs 010,\ \delta ({\rm J2000})=25\arcdeg 41\arcmin 45\farcs 500$. The Ka band observations are on 2021 October 23, while the Q band observations are from 2021 October to 2022 January (Oct. 20, 31; Nov. 1, 3, 7, 8; Dec 16; Jan. 3). The on-source observing time is $\sim 54$ min with the Ka band, while that with the Q band is 3.4 hr. The number of antennas was 27 with the Ka band and 28 with the Q band. The projected baseline length is 216 to 10090 m with the Ka band and 172 to 10175 m with the Q band. Any emission beyond  $6\farcs 6$ ($\sim 920$~au) in the Ka band or $4\farcs 5$ ($\sim 630$~au) in the Q band was resolved out by $\gtrsim 63\%$ with the antenna configuration \citep{1994ApJ...427..898W}. The bandpass calibrator was J0319+4130, the amplitude (flux) calibrator was 3C147, and the phase (gain) calibrator was J0403+2600.

The observations used 64 spectral windows, corving from $\sim 31$ to $\sim 38$~GHz with the Ka band and from $\sim 40$ to $\sim 48$~GHz with the Q band, and 4 polarization correlations (RR, RL, LR, and LL). Each spectral window covers 2 MHz$\times 64 =128$~MHz.

All the imaging procedure was carried out with Common Astronomical Software Applications (CASA), version 6.5.0.15. The visibilities were Fourier transformed and CLEANed with Briggs weighting and a robust parameter of 2.0. The resultant beam sizes and noise levels are describe in each figure in this paper. Furthermore, the coordinates are modified through the CASA task {\it fixplanet} so that the central position derived from the 2D Gaussian fitting through the CASA task {\it imfit} can be aligned between the two different bands. This modification is also applied to the other ALMA archival data so that every map has the same central position.
Even though we tried self-calibration for the VLA data using tasks in CASA ({\it tclean}, {\it gaincal}, and {\it applycal}), it did not increase the signal-to-noise ratio, and thus the VLA images in this paper are not self-calibrated.
While all maps are primary-beam corrected in this paper, the noise level were measured in emission-free areas before the primary-beam correction. The full width at half power of the primary beam is $\sim 63\arcsec$ with the Q band and $\sim 117\arcsec$ with the Ka band.

\subsection{ALMA Data Reprocessing} \label{sec:data}
We have reprocessed the archival ALMA data taken at Band 6 (2015.1.01415.S and 2018.1.00701.S) and Band 7 (2015.1.01549.S) that the $uv$-coverages overlapped with those of our VLA observations.
Most of these observations adopted quasar J0510+1800 as the absolute flux calibrator.
In our calibration procedures, the absolute flux densities of J0510+1800 were referenced from the ALMA Calibrator Grid Survey, the SMA Calibrator List \footnote{SMA Calibrator List: \url{http://sma1.sma.hawaii.edu/callist/callist.html}.}, and in some cases the archival SMA observations.
In 2014--2017, the flux density of J0510+1800 rapidly varied with time.
For certain epochs of ALMA observations, the rapid flux variability of J0510+1800 might have induced absolute flux inaccuracy. 
For this reason, these epochs were not used in our imaging process.
The details are commented on and summarized in Table \ref{tab:almadata} in Appendix \ref{appendix:almadata}.


\section{Results}\label{sec:result}
\subsection{Visibility profiles}\label{sec:vispro}
\begin{figure}[!htpb]
    \plotone{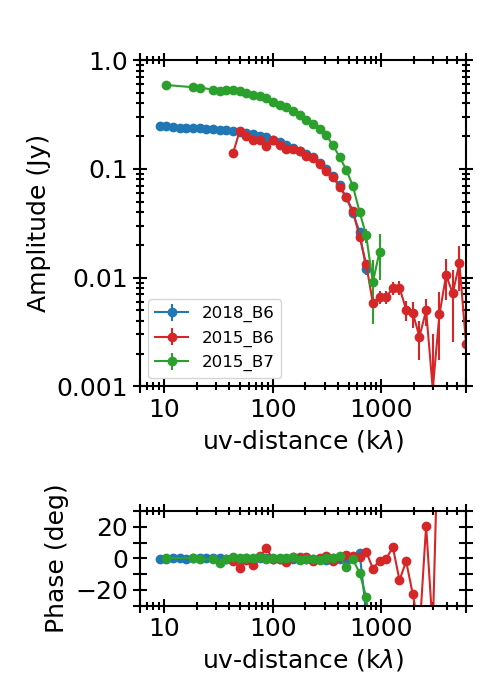}
    \caption{Comparison of 1D visibility profiles among the ALMA Band 6 and 7 data sets used in this paper. Each profile is derived by azimuthally averaging the visibilities on the 2D $uv$-plane after deprojection with the disk major axis of P.A.=$75\arcdeg$ and the disk inclination angle of $50\arcdeg$. The phase reference center is set to be the peak position of each data set.}
    \label{fig:vispro67}
\end{figure}

\begin{figure}[!htpb]
    \plotone{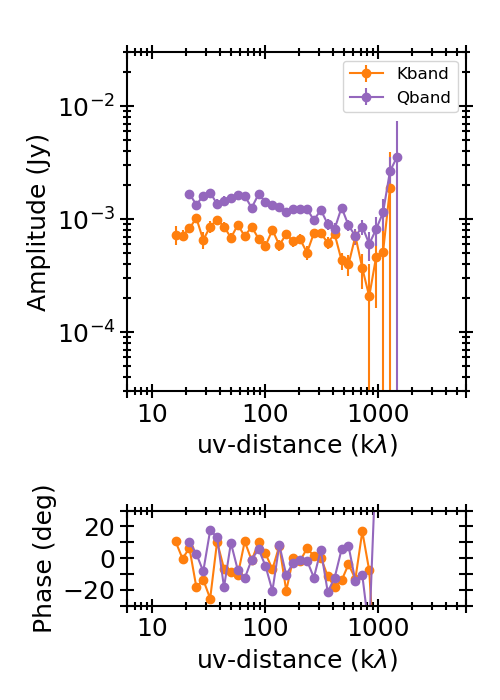}
    \caption{Same as Figure \ref{fig:vispro67} but for the VLA Ka and Q band data sets.}
    \label{fig:visproKQ}
\end{figure}

Figure \ref{fig:vispro67} shows comparison of 1D visibility profiles among the ALMA Band 6 and 7 data sets used in this paper, while Figure \ref{fig:visproKQ} shows that of the VLA Ka and Q band data sets. Each profile is derived from the visibilities on the 2D $uv$-plane. The $(u, v)$ coordinates of visibilities are deprojected assuming a geometrically thin disk structure, i.e., shrunk along the disk minor axis by a factor of $\cos i$, where $i$ is the disk inclination angle. The deprojection adopts the disk major axis direction of P.A.$=75\arcdeg$ and the inclination angle of $i=50\arcdeg$ \citep{2021ApJ...920...71A}. Then, the deprojected visilibities are azimuthally averaged and radially binned to 50 bins on a logarhismic grid from 6 to $6000~{\rm k}\lambda$ to derive the 1D profiles.

The ALMA amplitudes show that the two data sets of Band 6 (2015 and 2018) have amplitudes consistent with each other in the common $uv$-range. This figure also shows that the amplitude significantly decreases in $\gtrsim 200~{\rm k}\lambda$ in both ALMA bands. Assuming a Gaussian function, the half width at half maximum (HWHM) of $200~{\rm k}\lambda$ corresponds to the full width at half maximum (FWHM) of $\sim 0\farcs 5$ in the image domain, which is consistent with different past observations \citep{2015ApJ...812...27A}. In contrast, the VLA amplitudes show a shallower slope in both Ka and Q bands, even though the ALMA and VLA observations share most of the $uv$-range. This indicates that the VLA continuum emission arises from a central region. The HWHM appears $\sim 500~\mathrm{k}\lambda$, and this corresponds to the FWHM of $\sim 0\farcs 2$ of a Gaussian function in the image domain (see also Section \ref{sec:qk}).

\subsection{VLA Q and Ka bands}\label{sec:qk}
\begin{figure}[!htpb]
\gridline{
\fig{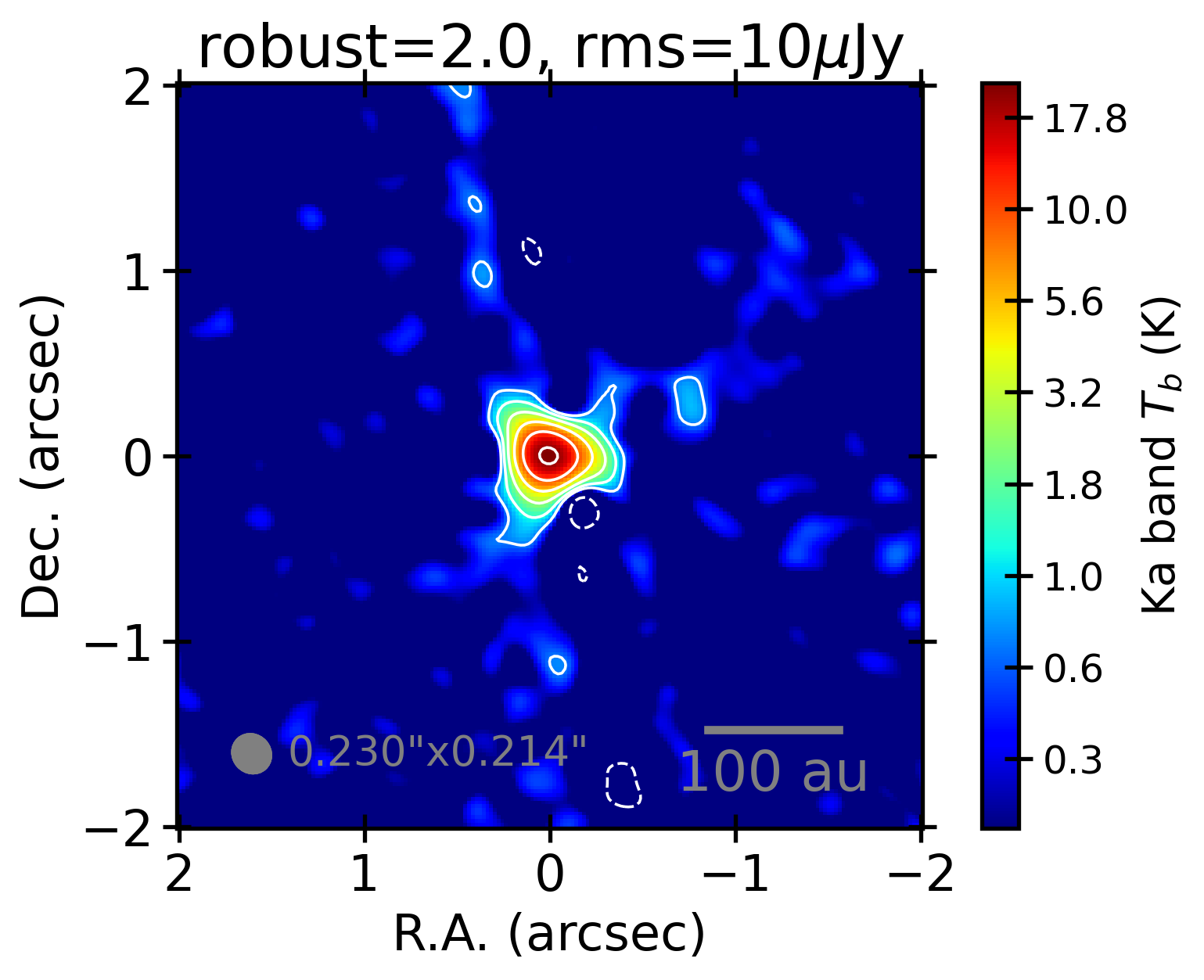}{0.45\textwidth}{(a) VLA Ka band (9 mm)}
}
\gridline{
\fig{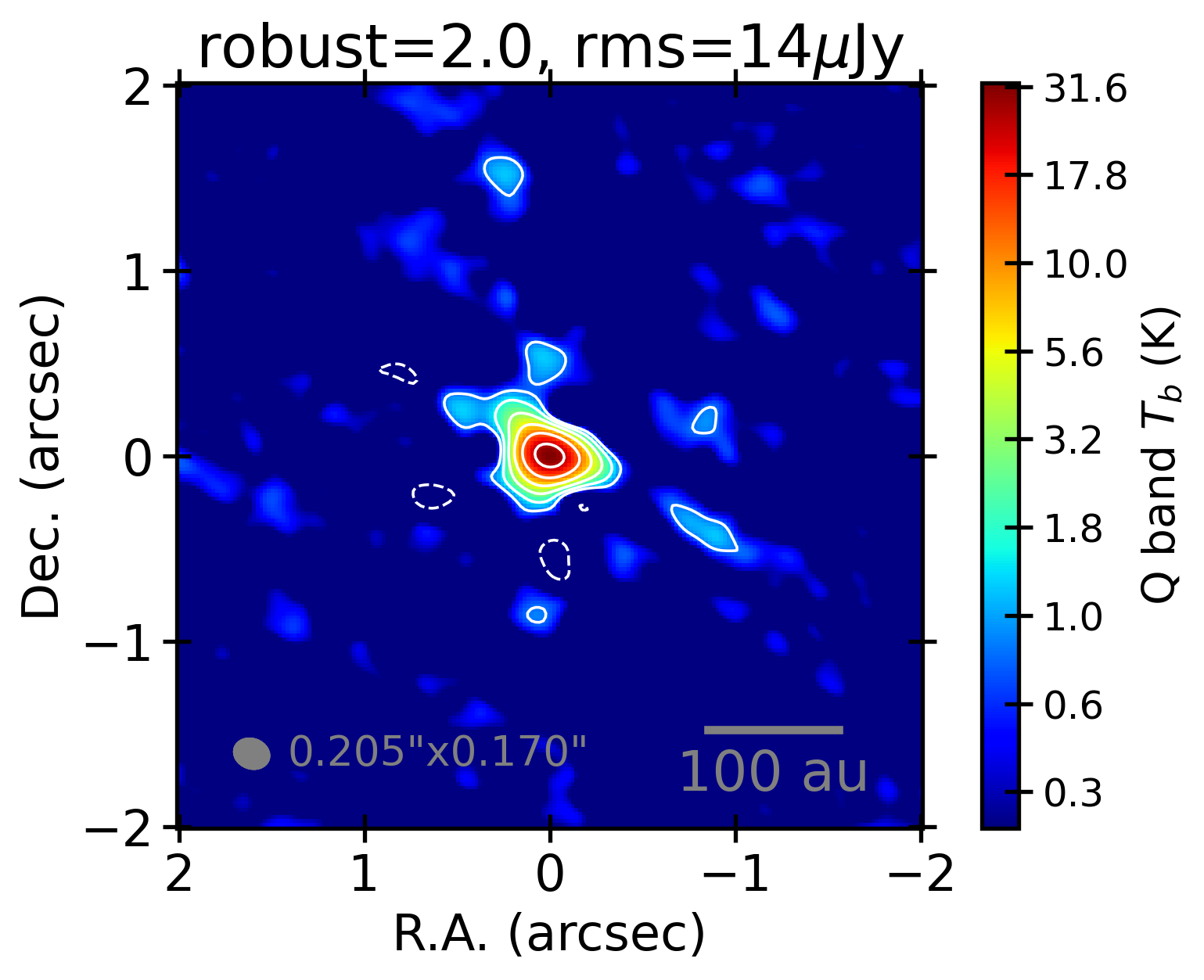}{0.45\textwidth}{(b) VLA Q and (7 mm)}
}
\caption{Continuum images in the VLA Ka and Q bands made with the robust parameter of 2.0, after the primary beam correction. The contour levels are in $3\sigma$ steps, where $1\sigma$ is (a) 10 and (b) 14~$\uJB$. The ellipse at the bottom left corner of each panel is the VLA synthesized beam.
\label{fig:qk}}
\end{figure}

Figure \ref{fig:qk} shows the continuum emission observed with our VLA observations in Ka (9 mm) and Q (7 mm) bands, after the primary beam correction. The abscissa and ordinate are the relative coordinates from the Gaussian center (Section \ref{sec:obs}). Both images show a compact component at the center and marginally detected components in the north. The northern component traces the eastern side of the blueshifted outflow lobe, which is stronger than the western side of the lobe in the $^{12}$CO $J=2-1$ line \citep{2021ApJ...920...71A}.
The Ka band image shows the peak intensity of $1.1~\mJB$ and the total flux density of 1.8~mJy, while those of the Q band image are $1.8~\mJB$ and 3.0~mJy, respectively.
The 2D Gaussian fitting provides the deconvolved size of $0\farcs 178\pm 0\farcs 008\times 0\farcs 114\pm 0\farcs 009$ (P.A.$=89\arcdeg \pm 6\arcdeg$) with the Ka band image and $0\farcs 179\pm 0\farcs 005\times 0\farcs 106\pm 0\farcs 004$ (P.A.$=75\arcdeg \pm 2\arcdeg$) with the Q band image.


\subsection{Spectral index map}\label{sec:spix}
\begin{figure*}[!htpb]
\gridline{
\fig{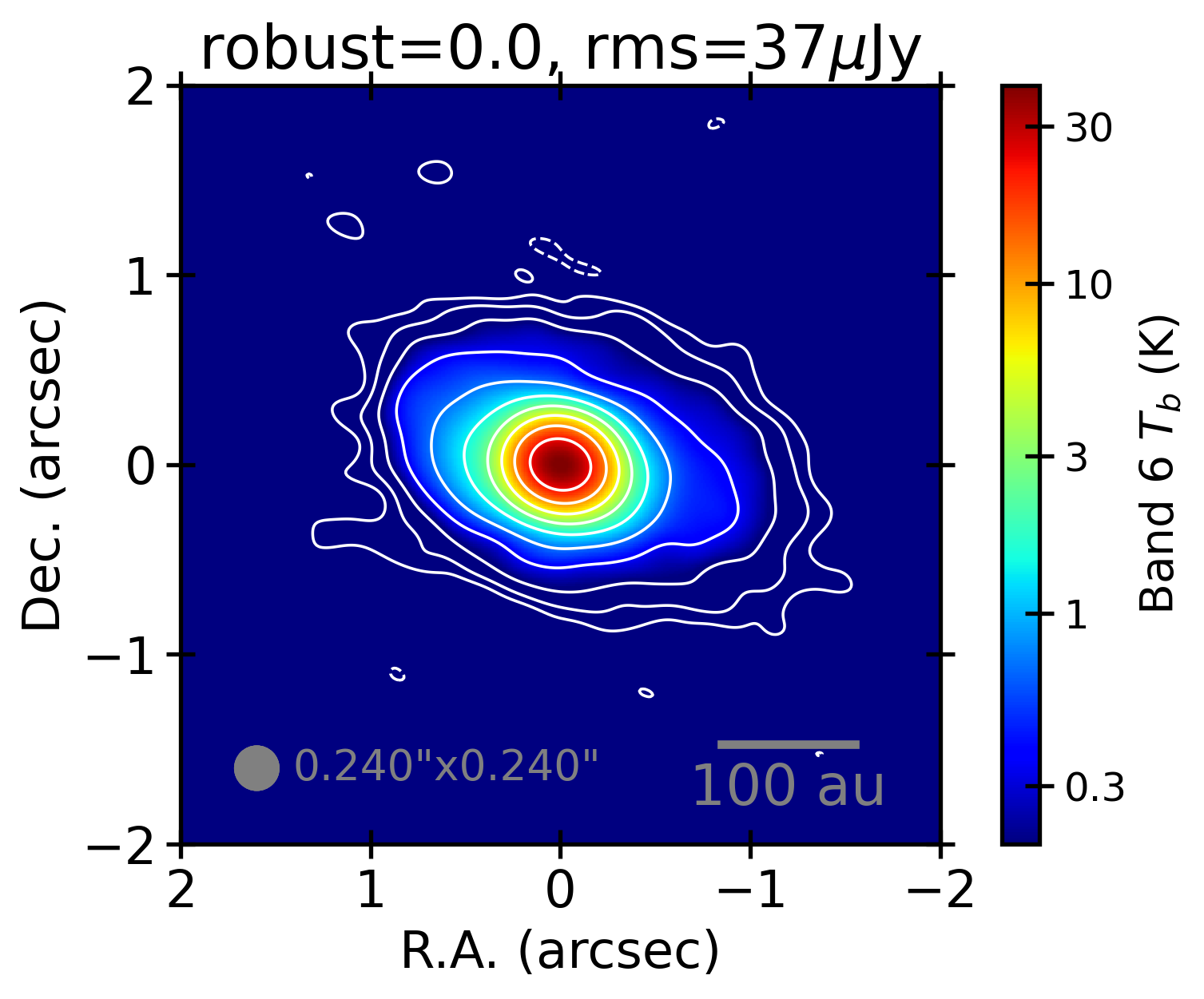}{0.32\textwidth}{(a) ALMA Band 6 (1.3 mm)}
\fig{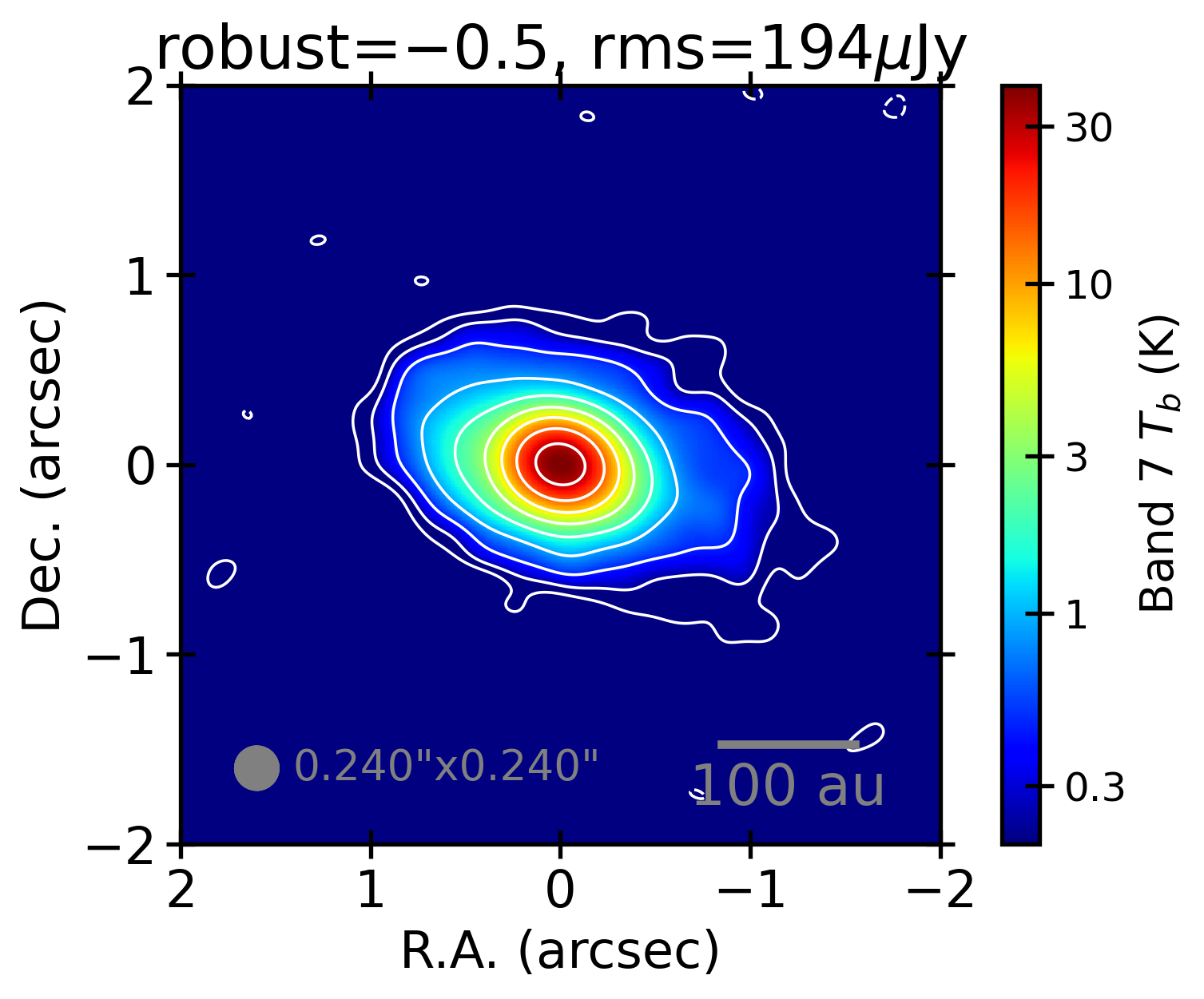}{0.32\textwidth}{(b) ALMA Band 7 (0.9 mm)}
\fig{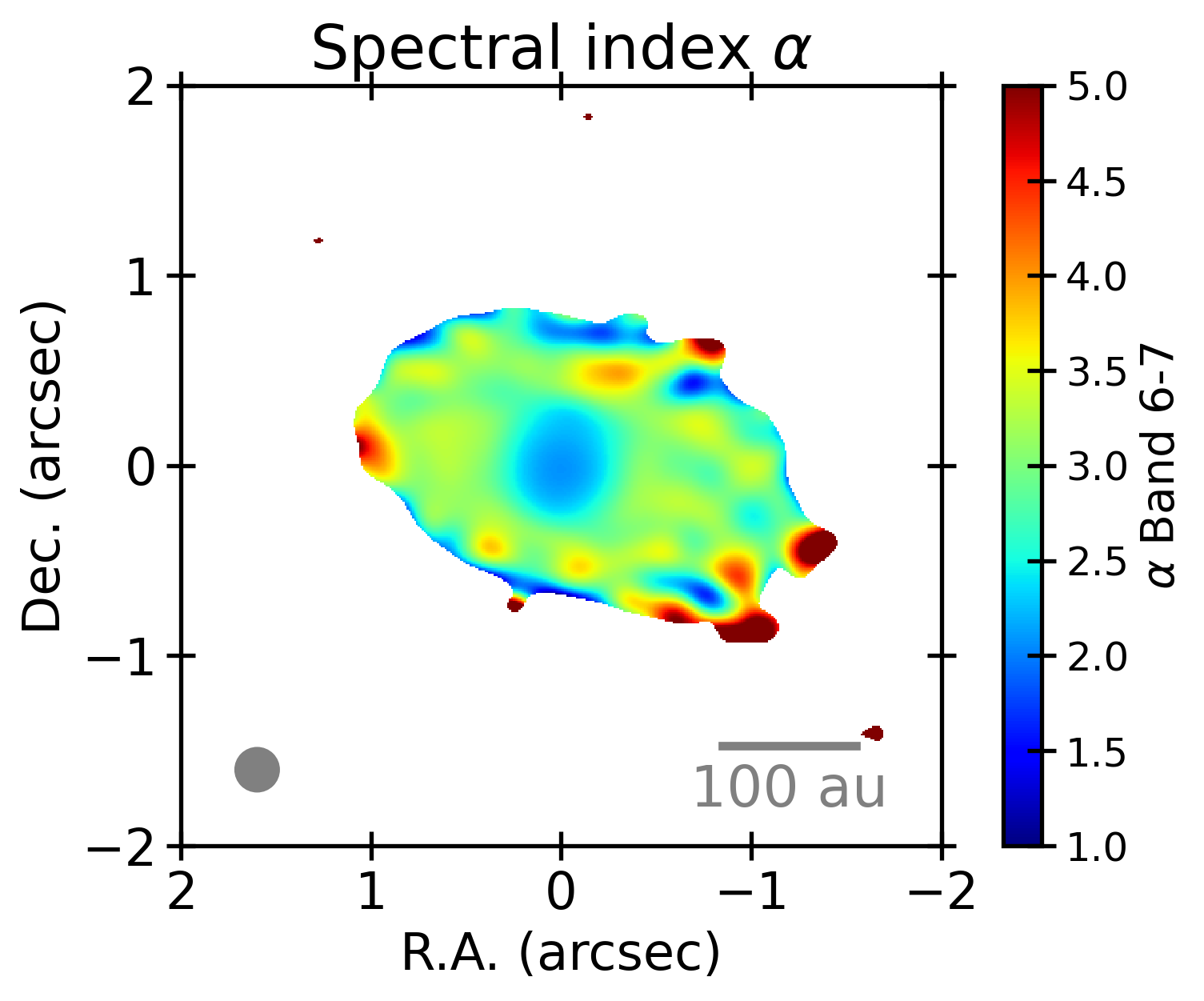}{0.32\textwidth}{(c) Spectral index}
}
\caption{Continuum images in the ALMA Band 6 and 7 made with the same $uv$ range of $20-1000~{\rm k}\lambda$, and the spectral index map made of the two images (panel c). Panels a and b are primary-beam corrected. Contour levels are in the $3\sigma$ steps, where $1\sigma$ is (a) 0.037 and (b) $0.19~\mJB$. The spectral index map is masked where the Band 7 continuum emission is below $3\sigma$. The ellipse at the bottom left corner of each panel is the synthesized beam common for all panels ($0\farcs 24 \times 0\farcs 24$). The common beam is achieved by $uv$-tapering and spatial smoothing after a similar beam is selected by the robust parameter.
\label{fig:alpha67}}
\end{figure*}
The combinations of the newly observed VLA Ka and Q band images, as well as the ALMA Band 6 and 7 images from the archive, enable us to measure the spectral index $\alpha = d\ln F_\nu /d\ln \nu$, at each position, where $F_\nu$ and $\nu$ are the flux density and frequency, respectively. To calculate the index $\alpha$, the two images must be made from visibilities with the same $uv$ range and have the same synthesized beam size. Hence, in this subsection, we limit the $uv$ range to a common range of $20-1000~{\rm k}\lambda$ to make all the four images. Then, the robust parameter for the CASA task {\it tclean} is selected to obtain a synthesized beam size close but smaller than the target resolution $0\farcs 24$: 0.0 for ALMA Band 6, $-0.5$ for ALMA Band 7, 2.0 for VLA Ka band, and 2.0 for VLA Q band. The synthesized beam size is further tuned to be closer to, but still smaller than, the target resolution by $uv$-tapering: $0\farcs 167\times 0\farcs 067$ ($-79.4\arcdeg$) for ALMA Band 6, $0\farcs 218\times 0\farcs 065$ ($-103.5\arcdeg$) for ALMA Band 7, $0\farcs 076\times 0\farcs 033$ ($133.3\arcdeg$) for VLA Ka band, and $0\farcs 153\times 0\farcs 066$ ($162.4\arcdeg$) for VLA Q band. Finally the synthesized beam size is set to be exactly the target resolution by spatial smoothing through the CASA task {\it imsmooth}. Using these images at the common resolution, the $\alpha$ map is made through the CASA task {\it immath}.

Figure \ref{fig:alpha67} shows the ALMA Band 6 and 7 images used to make the $\alpha$ map and the resultant $\alpha$ map. The $\alpha$ values are masked where the Band 7 emission is weaker than the $3\sigma$ level. The ALMA Band 6-7 $\alpha$ is closer to 2 at the central $\sim 40$~au, while it is $\sim 3.3$ in the outer region (Figure \ref{fig:alpha67}c). The index $\alpha \sim 2$ can be confirmed as the same brightness temperature in Band 6 nad 7 $\sim 32$~K in the central region (Figure \ref{fig:alpha67}ab). This index indicates that the continuum emission is optically thick, which is consistent with the absorption in the C$^{18}$O and $^{13}$CO $J=2-1$ lines against the strong continuum emission \citep{Harsono2018NatAs...2..646H}.

At the region where the Band 6 and 7 brightness temperatures are $\lesssim 10$~K, the Band 6-7 $\alpha$ is consistent with a spatially uniform value of $\sim 3.3$. 
A common way to interpret the $\alpha$ index is to express it as a function of the optical depth $\tau$ and the dust opacity index $\beta$ as follows:
\begin{eqnarray}
\alpha = \frac{d\ln (B_\nu (T_d)(1-e^{-\tau_\nu}))}{d\ln \nu}
\nonumber\\
=\frac{d\ln B_\nu(T_d)}{d\ln \nu} + \frac{\tau_\nu}{e^{\tau_\nu}-1}\beta,
\label{eq:beta}
\end{eqnarray}
where $B_\nu$, $T_d$, and $\nu$ are the Planck function, dust temperature, and frequency, respectively. 
For example, for an assumed dust temperature $T_d\sim 20-30$~K, $d\ln B_\nu /d\ln \nu \sim 1.7$ at the middle frequency of our Band 6 and 7 observations, 277~GHz (1.1~mm).
In this case, if the Band 6 and 7 observations are (marginally) optically thin ($\tau_\nu < 1$), which is likely the case given the resolved $\lesssim 10$~K dust brightness temperatures, the observed $\alpha$ value would imply that $\beta$ is approximately between 1.6 ($\tau _\nu \ll 1$) and 2.7 ($\tau _\nu =1$).
This $\beta$ value is larger than that of the interstellar dust, which 
may indicate that the maximum grain size $a_{\rm max}$ is on the order of 100s~$\micron$ \citep[$\beta$ around $\lambda=$1.3 mm;][]{2019MNRAS.486.3907P}. 
In contrast, $a_{\rm max}>$1~mm produces the index of $\beta \lesssim 1.0$ \citep{2019MNRAS.486.3907P}.

\begin{figure*}[!htpb]
\gridline{
\fig{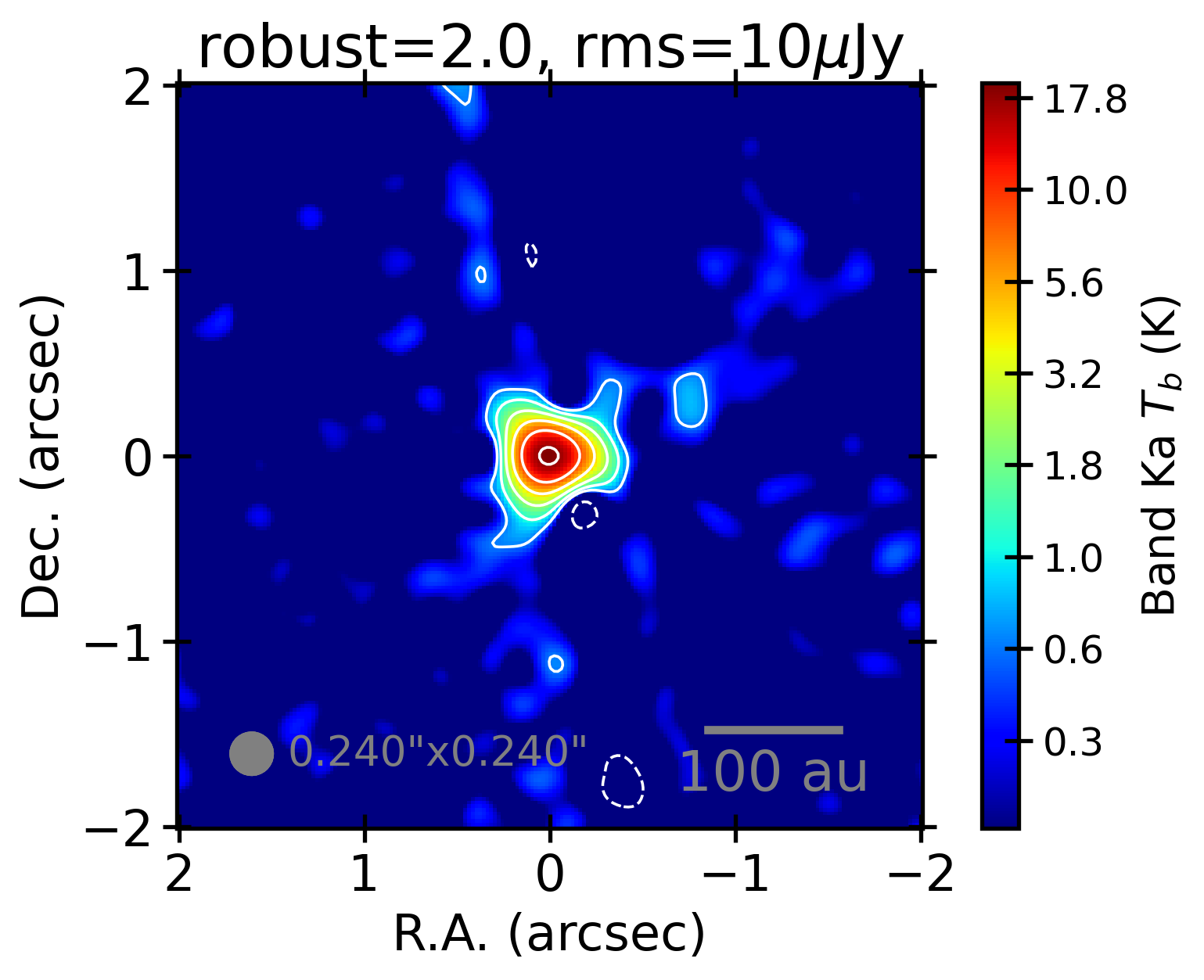}{0.32\textwidth}{(a) VLA Ka band (9 mm)}
\fig{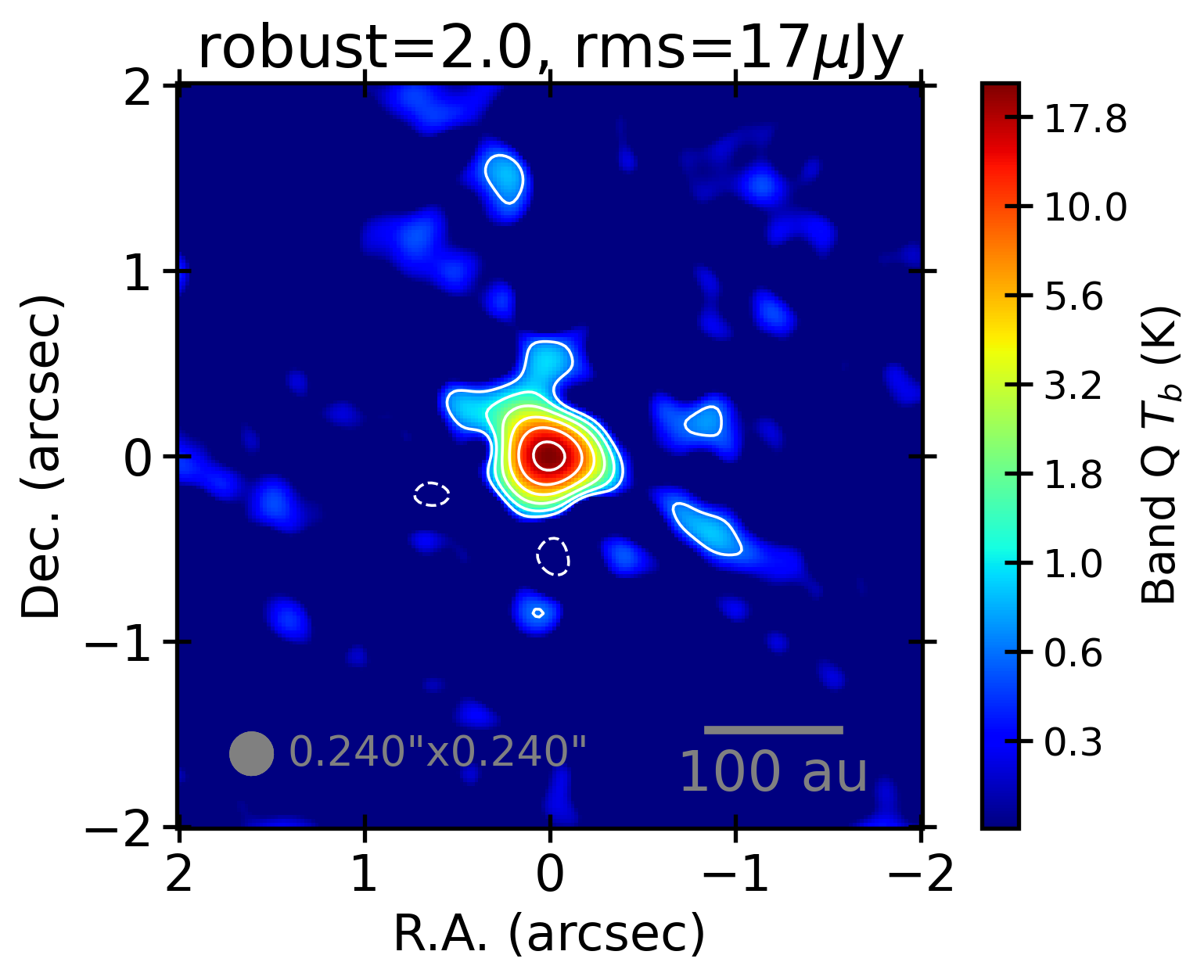}{0.32\textwidth}{(b) VLA Q band (7 mm)}
\fig{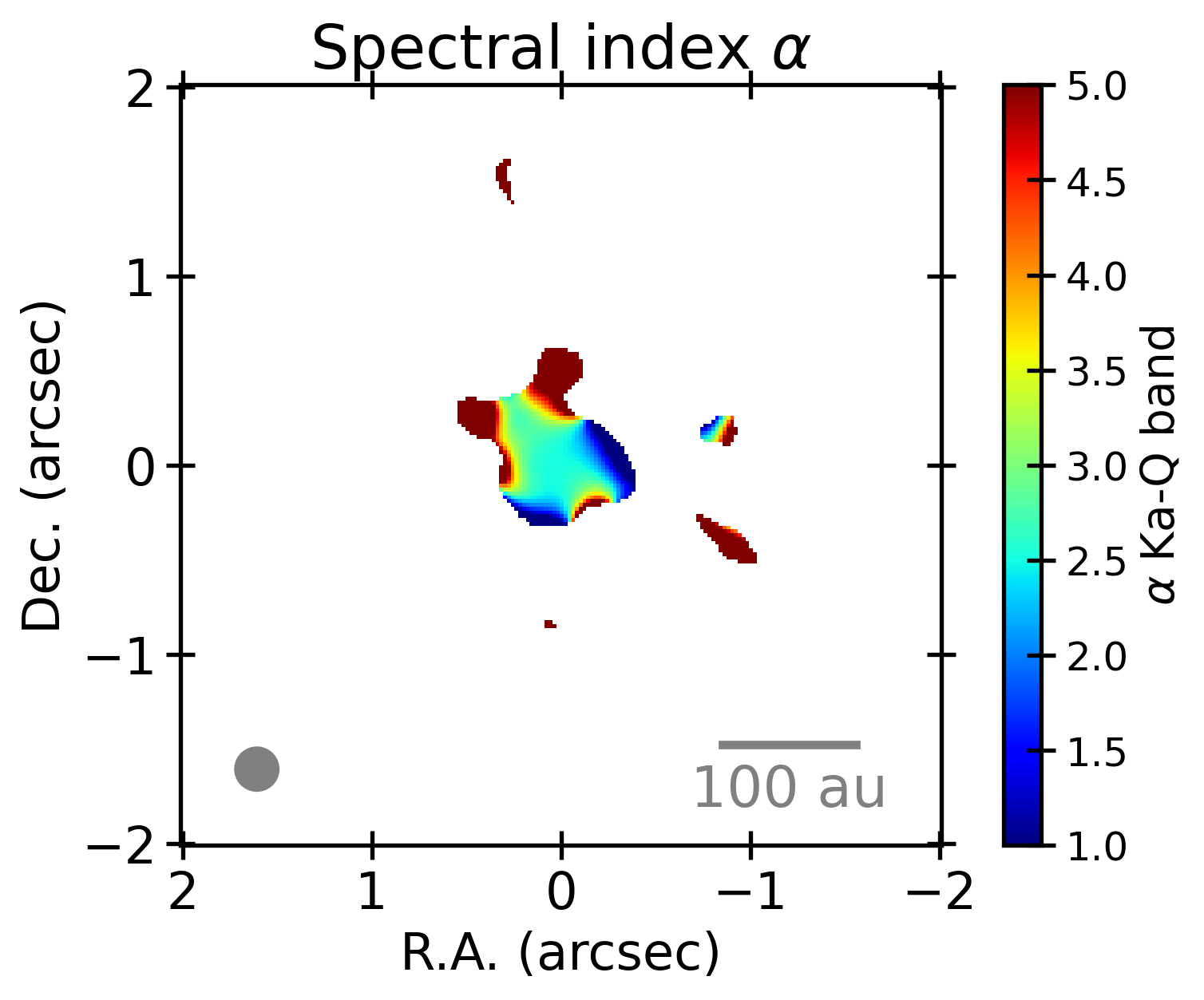}{0.32\textwidth}{(c) Spectral index}
}
\caption{Same as Figure \ref{fig:alpha67} but for the VLA (a) Ka and (b) Q band images, and (c) the spectral index map made of the two images. The $1\sigma$ level is (a) 10 and (b) $17~\uJB$. The spectral index map is masked where the Q band continuum emission is below $3\sigma$.
\label{fig:alphakq}}
\end{figure*}

Figure \ref{fig:alphakq} shows the VLA Ka and Q band images that were used to make the $\alpha$ map and the resultant $\alpha$ map. 
The $\alpha$ values are masked where the Q band emission is weaker than the $3\sigma$ level. 
The VLA Ka-Q band $\alpha$ is measured only at the central $\sim 40$~au region at $\alpha \sim 2.5$. 
Because the brightness temperature in the Ka and Q images is lower than in the ALMA Band 6 and 7 images, the continuum emission is marginally optically thin at Ka and Q bands, if the beam filling factor is unity. 
The index of the Planck function is $d\ln B_\nu /d\ln \nu \sim 2.0$ at the middle frequency of the our Ka and Q band observations, 31~GHz (0.95~cm), and $T_d> 30$~K. 
If the optical depth is $\tau_\nu \lesssim 1$, Equation (\ref{eq:beta}) provides $\beta$ between 0.5 ($\tau_\nu \ll 1$) and 0.9 ($\tau _\nu =1$), which is much smaller than the $\beta$ index measured between ALMA Band 6 and 7. 
This face value of $\beta$ may imply that the innermost region of the TMC-1A disk has a larger $a_{\rm max}$ value, although the exact $a_{\rm max}$ value needs to be examined by the SED fittings that take the beam filling factor and dust scattering opacities (e.g., \citealt{Liu2019ApJL}) into considerations.
Such SED analyses are introduced in Section \ref{sec:sed}.


\section{Analysis}
\subsection{Spectral energy distributions} \label{sec:sed}
We produced slices of the dust continuum intensity along the major axis of the TMC~1A disk at the observed frequency bands.
To avoid cluttering presentations, we only discuss the spectral energy distribution (SED) models for certain selected offset positions.
In addition, instead of sampling the slice at a regular spatial interval, we present more samples on the small spatial scales due to the steeper variations of dust properties (temperature, column density, grain sizes) on smaller spatial scales.
Dust properties in the regions between our selected spatial samples can be understood in an interpolated sense. 

Our strategy follows that of \citet{Liu2019ApJ...884...97L,Liu2021ApJ...923..270L} to model the observed flux densities as the contributions of various dust emission components.
The exact formulation is as follows:
\begin{equation}\label{eqn:multicomponent}
    F_{\nu} = \sum\limits_{i} F_{\nu}^{i} e^{-\sum\limits_{j}\tau^{i,j}_{\nu}}, 
\end{equation}
where $F_{\nu}^{i}$ is the flux density of the dust component $i$ and $\tau^{i,j}_{\nu}$ is the optical depth of the emission component $j$ that obscures the emission component $i$. 

To take into account the effects of dust self-scattering (e.g., \citealt{Liu2019ApJL}), the flux density of each dust component is evaluated through Equations (11)--(19) in \citet{Birnstiel2018ApJ...869L..45B}.
Limited by the angular resolution, the spatially resolved regions are outside of the water snowline ($\sim$150 K; e.g., \citealt{Pollack1994ApJ...421..615P}).
Optical constants are, therefore, quoted from the default (i.e., water-ice-coated) DSHARP dust (\citealt{Birnstiel2018ApJ...869L..45B}).
To produce the grain-size weighted dust opacities, we assumed a power-law dust ($a^{q}$) grain size ($a$) distribution $n(a)$ between the minimum and maximum grain sizes ($a_{\mbox{\scriptsize min}}$, $a_{\mbox{\scriptsize max}}$), adopting a nominal $a_{\mbox{\scriptsize min}}=10^{-4}$~mm.
This modeling tries $q=-2.5$ and $q=-3.5$, which represent the drift-limited and fragmentation limited grain growth, respectively (\citealt{Birnstiel2012A&A...539A.148B}). 
As a summary, the free parameters for each dust components are dust temperature ($T_{\mbox{\scriptsize dust}}$), mass surface density ($\Sigma_{\rm dust}$), maximum grain size $a_{\mbox{\scriptsize max}}$, and   the beam filling factor $f_{\mbox{\scriptsize beam}}$ when fitting with more than one dust components.

For each selected position, we produced (1) the SED model that assumes uniform dust properties in a line-of-sight (i.e., there is only one dust component), and (2) the SED model with two dust components in a line-of-sight: a surface and a mid-plane dust components.
The results of our SED modelings for cases (1) and (2) are discussed in Section \ref{sub:1dust} and \ref{sub:2dust}, respectively.
Case (2) needs to be considered because using only one dust component cannot reproduce the SEDs at every offset position simultaneously.
On the other hand, the number of our independent measurements does not permit assuming more than two dust components in a line-of-sight. 

In case (1), the beam filling factor ($f_{\mbox{\scriptsize beam}}$) is 1.0 everywhere.
Due to the degeneracy between dust temperature and $f_{\mbox{\scriptsize beam}}$, taking $f_{\mbox{\scriptsize beam}}$ as a free parameter does not yield the better reliable fittings.

In case (2), $f_{\mbox{\scriptsize beam}}$ of the surface component is 1 while $f_{\mbox{\scriptsize beam}}$ of the mid-plane component is a free parameter.
An additional assumption is that the mid-plane component is obscured by the surface component on the near side, and the surface component on the far side is obscured by both the mid-plane component (taking into account the factor $f_{\mbox{\scriptsize beam}}$) and the surface component in the near side.
In case (2), the assumptions of $f_{\mbox{\scriptsize beam}}$ are hard to avoid due to the limited number of observed frequency bands.
Nevertheless, we argued that the assumed 1.0 filling factor at the surface is realistic given the high gas and dust column density in the Class 0/I sources.
Since Class 0/I sources may form substructures in the high-density mid-plane (e.g., \citealt{Sheehan2018ApJ...857...18S,Nakatani2020ApJ...895L...2N,Sheehan2020ApJ...902..141S,SeguraCox2020Natur.586..228S}), given the synthesized beam sizes of the utilized VLA and ALMA observations, it is possible that the $f_{\mbox{\scriptsize beam}}$ is, in general, less than 1.0 in the mid-plane. 

It turned out that the assumptions of $q=-2.5$ and $q=-3.5$ led to qualitatively similar fittings.
When fitting with more than one dust components in a line-of-sight, we assumed that the surface and the mid-plane components have the same $q$ and $a_{\mbox{\scriptsize max}}$ values to limit the overall number of free parameters.
This assumption is not necessarily realistic.
Since the large dust particles may preferentially settle to the disk mid-plane, the surface and mid-plane components may have different $a_{\mbox{\scriptsize max}}$ values. 
Relaxing these assumptions in principle can yield better fittings, which are however not necessarily scientically meaningful since it will make the fitting more under-constrained.

Flux densities at centimeter wavelengths could be contributed by the free-free emission from a protostellar jet \citep{1998AJ....116.2953A, 2014ApJ...780..155L, 2015ApJ...801...91D, 2018ApJ...852...18T, 2019A&A...631A..58C}. We assess the importance of free-free emission based on the archival VLA observations at lower frequencies and found that free-free emission is likely negligible as compared to dust emission in the specific case of TMC1A. The evaluation of the free-free emission is discussed in Section \ref{sec:free} in more detail.

\subsection{One dust component}\label{sub:1dust}
The free parameters are optimized using the Markov chain Monte Carlo (MCMC) method through the {\tt emcee} (\citealt{Foreman-Mackey2013PASP}) package. The number of walkers, steps, and burn-in steps are 60, 1500, and 500, respectively.
The fittings occured to be degenerated: one branch of solutions have $a_{\mbox{\scriptsize max}}>1$ mm in general (hereafter the grown dust branch), while the other branch of solutions have $a_{\mbox{\scriptsize max}}\ll1$ mm (hereafter the small dust branch).
The prior probability functions are uniform but bounded at $a_{\mbox{\scriptsize max}}\gtrless 0.5$ mm to focus the fittings on one branch at a time.

\begin{figure*}[!htpb]
    \hspace{1.5cm}
    \begin{tabular}{c}
         \includegraphics[width=15cm]{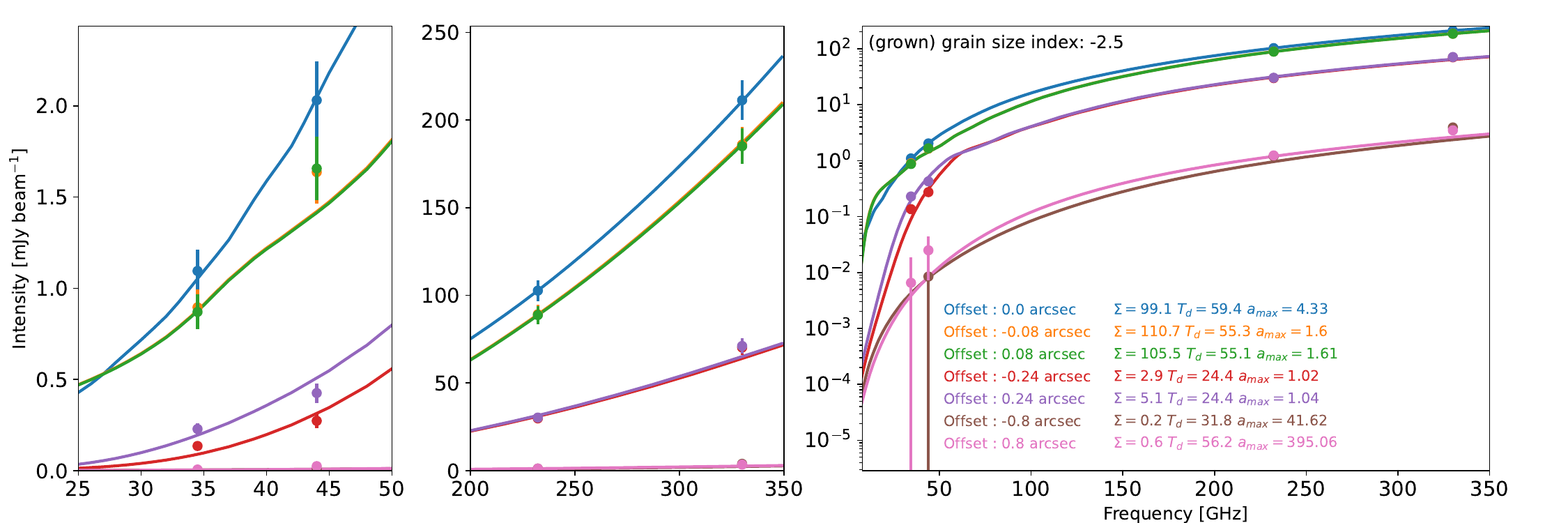} \\
         \includegraphics[width=15cm]{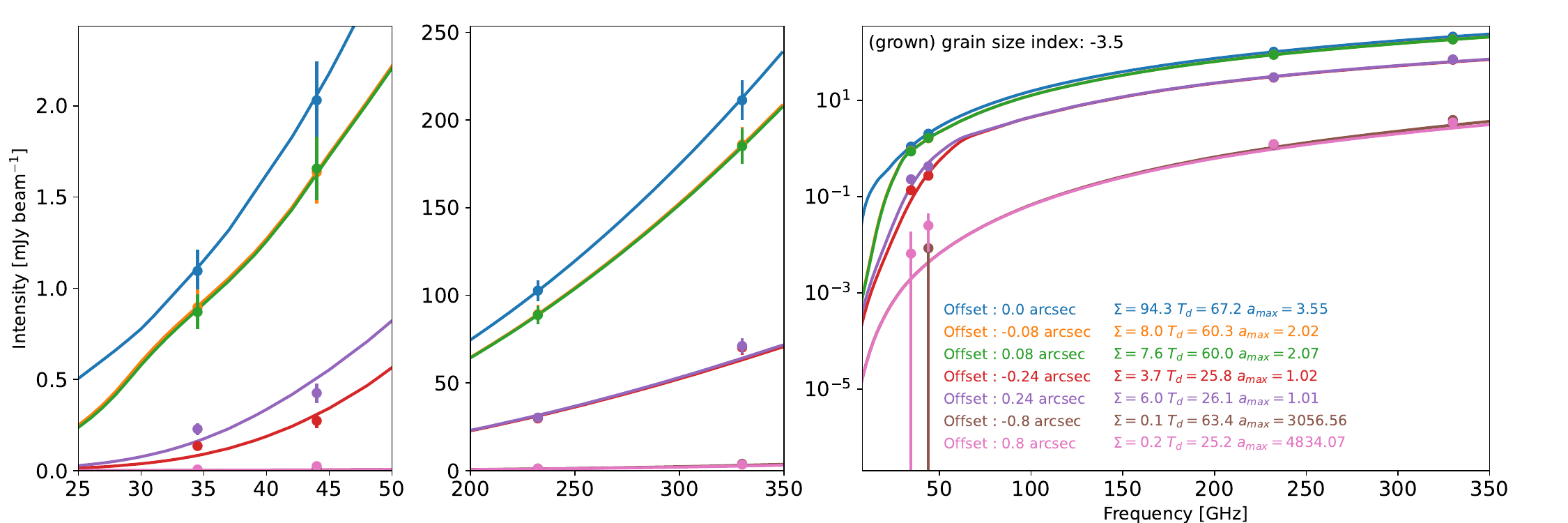} \\
         \includegraphics[width=15cm]{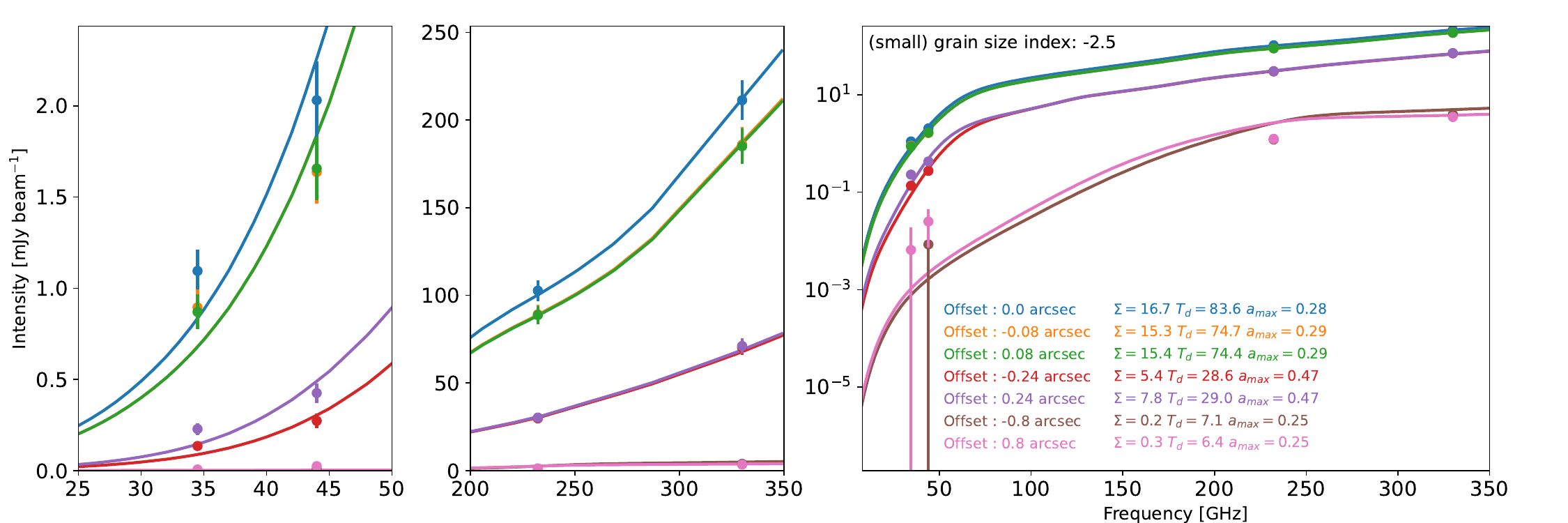} \\
         \includegraphics[width=15cm]{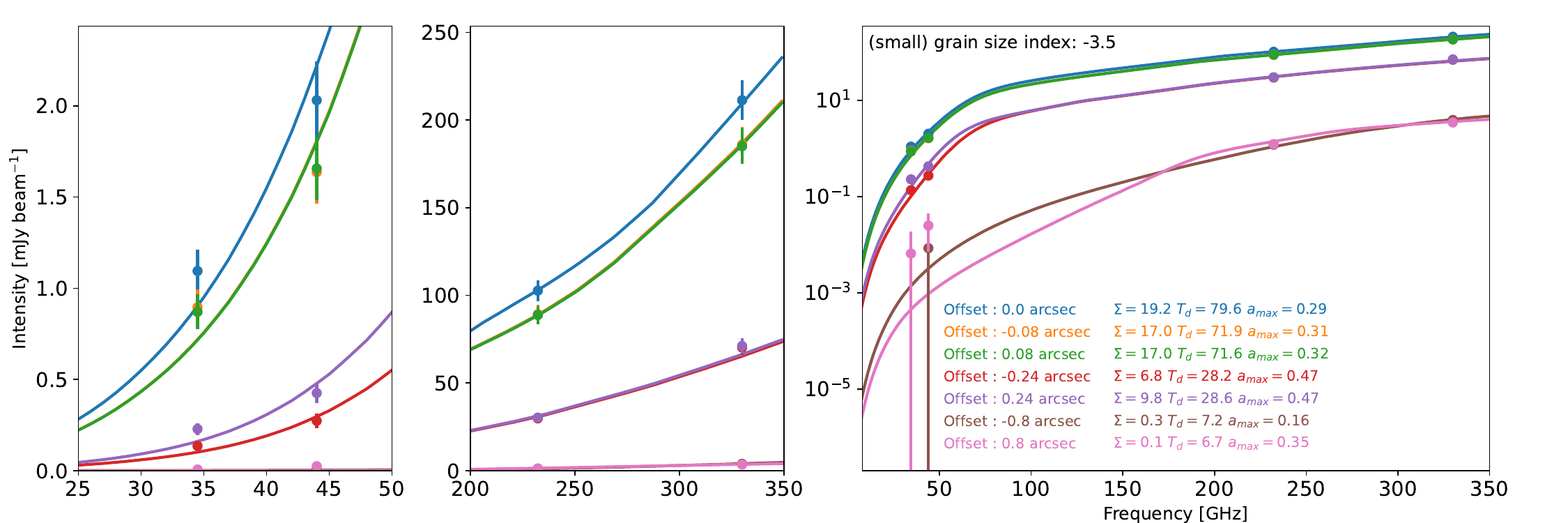} \\
    \end{tabular}
    \caption{Spectral energy distribution fitting assuming a single DSHARP dust component. The observed intensities are shown by dots that are color-coded by the offset positions. From top to bottom rows, the solid lines show the SED models assuming (1) grown dust branch with $q=-2.5$, (2) grown dust branch with $q=-3.5$, (3) small dust branch with $q=-2.5$, and (4) small dust branch with $q=-3.5$.}
    \label{fig:singledsharp_sed}
\end{figure*}

Figure \ref{fig:singledsharp_sed} shows the grown and small dust branch solutions for the cases with grain size power-law index  $q=-2.5$ and $q=-3.5$.
With $q=-2.5$, in both the grown and small dust branches, the observed fluxes can be fitted simultaneously at all four frequency bands only at the central location (i.e., offset$=$0$''$). 
Therefore, the single-compoment dust model with $q=-2.5$ is disfavored.
With $q=-3.5$, the grown dust branch can simultaneously fit the measurements at all four frequency bands only in the offset range of $\pm$0\farcs08 ($\sim$11 au).
The small dust branch can only simultaneously fit the measurements at all four frequency bands at the central location.
In our best-fit models, at the central location, the $a_{\mbox{\scriptsize max}}$ is $\sim$2--4 mm in the grown dust branch, while $a_{\mbox{\scriptsize max}}$ is $\sim$0.3 mm in the small dust branch.

At larger offsets, the best-fit models yield spectral indices that are steeper than the observed value between VLA Q and Ka bands.
We note that in these cases, the parameter $a_{\mbox{\scriptsize max}}$ did not reliably converge in the MCMC fitting, in particular, when the offsets are comparably larger than 0\farcs8 (112 au): the high $a_{\mbox{\scriptsize max}}$ values in the grown dust branch merely reflect that the prior function is not upper-bounded, while the modest $a_{\mbox{\scriptsize max}}$ values in the small dust branch is a result that the $a_{\mbox{\scriptsize max}}$ values of the MCMC samplers were randomly walking in the range of 0--0.5 mm.

These fitting results imply that only one dust component in a line-of-sight cannot sufficiently explain the observations. The good models we obtained at the central location may be coincidence, i.e., they can be biased due to the oversimplified assumption of thermal and density structures. 

\clearpage

\subsection{Separating dust mid-plane and surface}\label{sub:2dust}
\begin{figure}[!htpb]
    \centering
    \includegraphics[width=8cm]{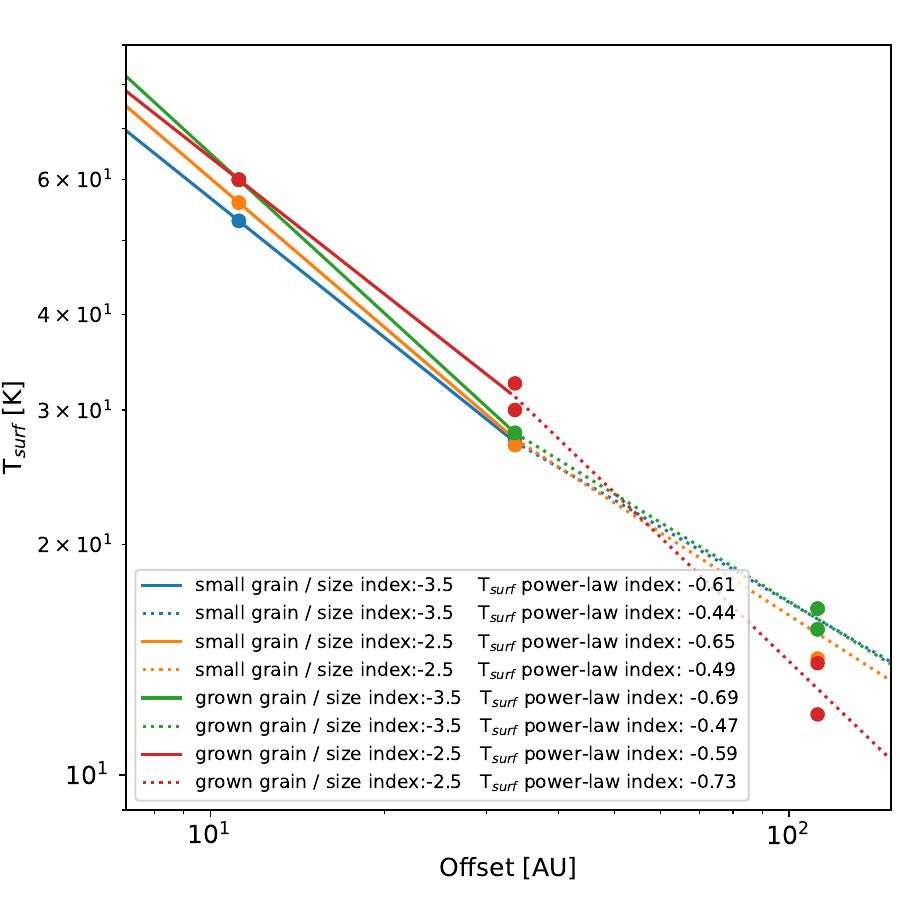}
    \caption{Derived surface dust temperature profiles (see Section \ref{sub:2dust}).}
    \label{fig:Tprofile}
\end{figure}

We interactively search for the parameters that are consistent with the data.
For each line-of-sight, in spite that there are more free parameters than independent measurements, the temperature of the surface component can still be constrained to a certain extent due to the modest optical depth (e.g., implied by the spectral indices between ALMA Band 6 and 7).
The dust scattering opacity is very sensitive to $a_{\mbox{\scriptsize max}}$, and the attenuation due to the dust scattering opacity can systematically bias the derivation of $T_{\mbox{\scriptsize dust}}$ (e.g., \citealt{Liu2019ApJL}).
Nevertheless, in our present case study, this effect is very modest except at the central location. 
In both grown and small dust branches, with $q=-2.5$ and $q=-3.5$, the radial surface dust-temperature profiles may be approximated by (broken) power-laws: it is either consistent with a $\sim-0.7$ power-law index, or is consistent with a $-0.6$ - $-0.7$ power-law index at the inner $\sim$30 au radii and a $-0.4$ - $-0.5$ power-law index at the outer $\sim$30--100 au radii (Figure \ref{fig:Tprofile}).

\begin{figure*}[!htpb]
    \hspace{1.5cm}
    \begin{tabular}{c}
         \includegraphics[width=15cm]{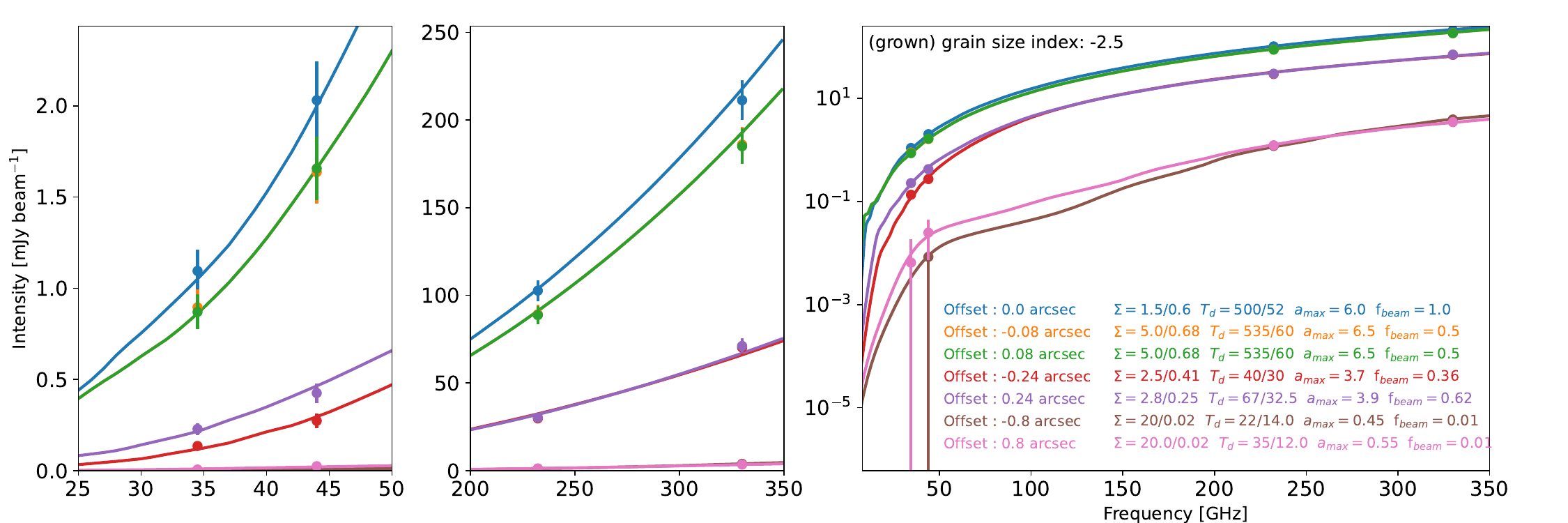} \\
         \includegraphics[width=15cm]{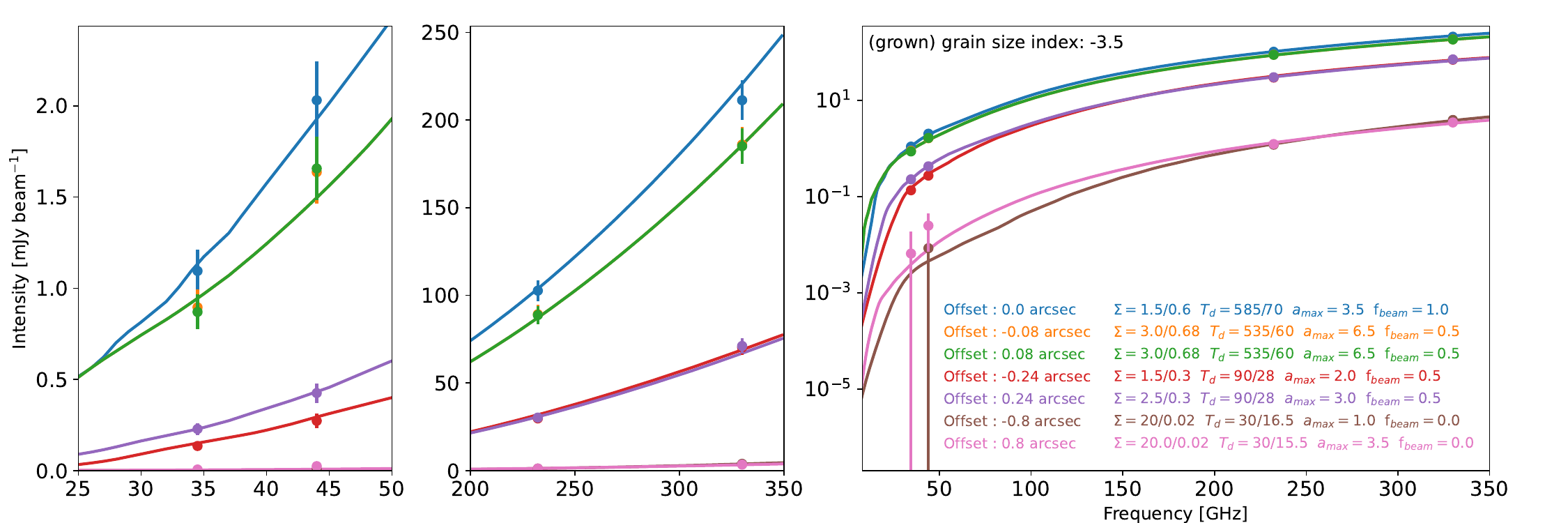} \\
         \includegraphics[width=15cm]{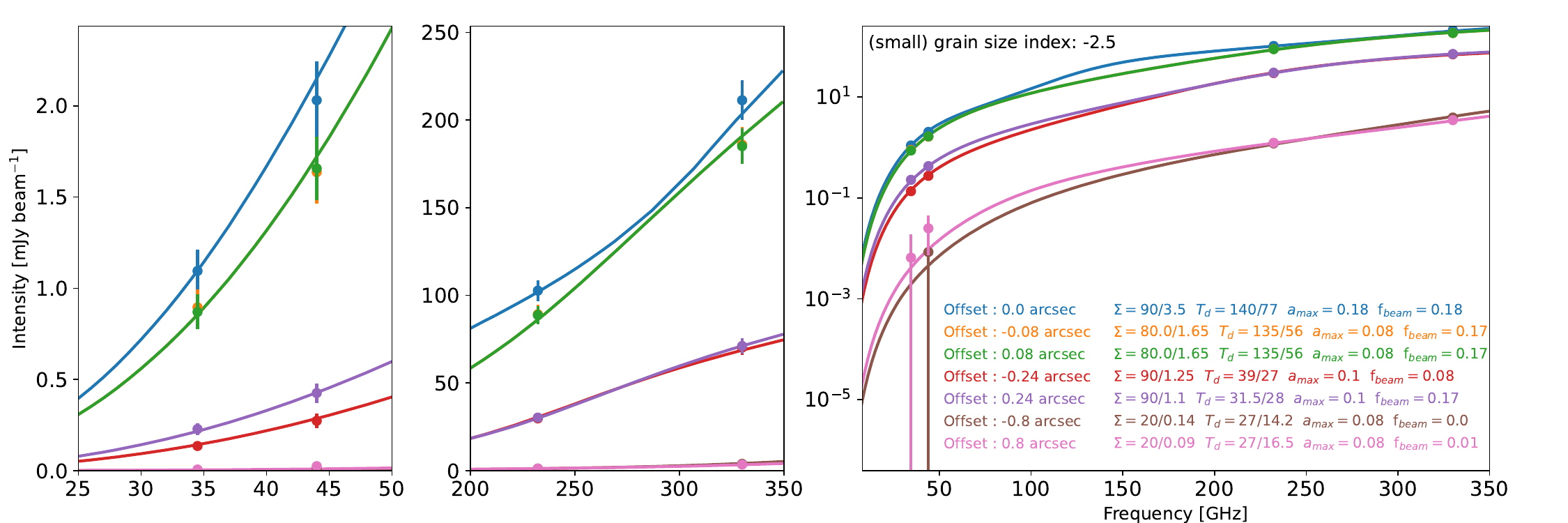} \\
         \includegraphics[width=15cm]{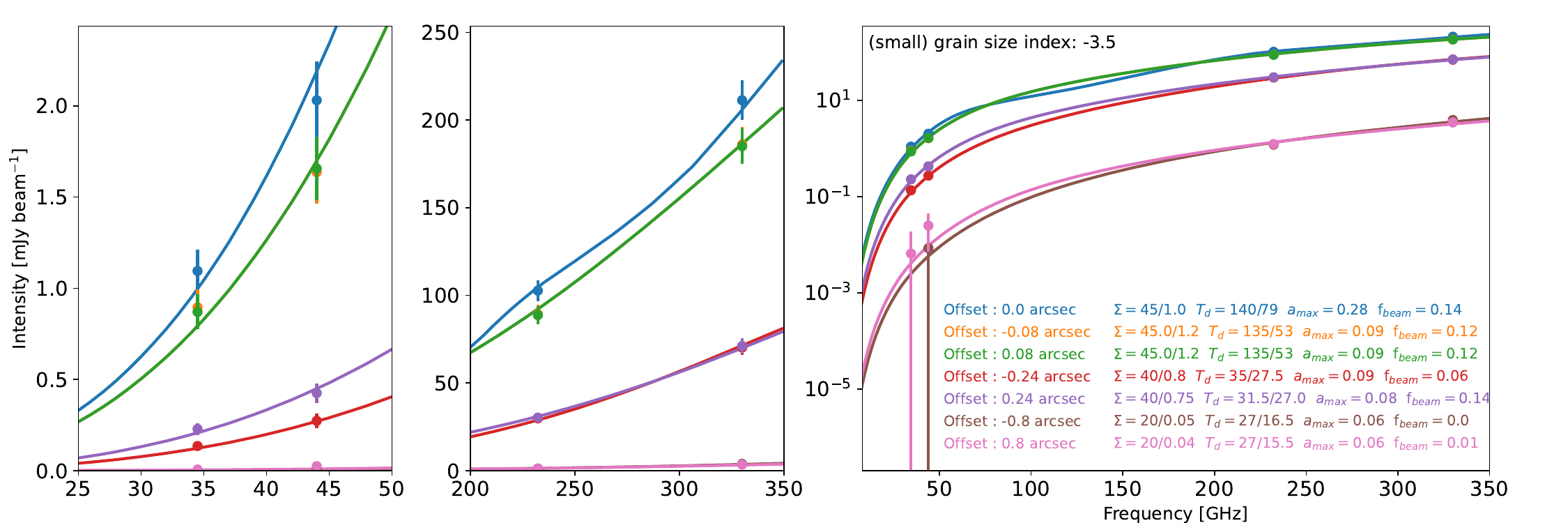} \\
    \end{tabular}
    \caption{Same as Figure \ref{fig:singledsharp_sed} but the SED models assume that there are mid-plane and surface dust components (see Section \ref{sub:2dust}).}
    \label{fig:twodsharp_sed}
\end{figure*}

Including a mid-plane dust component that is warmer than the surface component and has $f_{\mbox{\scriptsize beam}}<1$ can enhance the flux density at 29--48 GHz, as shown in Figure \ref{fig:twodsharp_sed}. The best-fit models in Figure \ref{fig:twodsharp_sed} have $a_{\mbox{\scriptsize max}}\sim 0.12$~mm in the small dust branch and $\sim 4$~mm in the grown dust branch as a mean value among the radii.
When a small $f_{\mbox{\scriptsize beam}}<1$ and a high $\Sigma_{\mbox{\scriptsize dust}}$ are used, the degree of freedom makes it possible to reproduce the shallow spectral indices observed between the VLA Q and Ka bands.
In the grown dust branch, the relatively high $T_{\mbox{\scriptsize dust}}$ and $f_{\mbox{\scriptsize beam}}$ values are needed for the mid-plane component because the emission from the mid-plane component is strongly attenuated by the scattering opacity of the surface component. The $T_{\mbox{\scriptsize dust}}$ and $f_{\mbox{\scriptsize beam}}$ values for the mid-plane component can be lowered if the grown dust is only present in the disk mid-plane (i.e., the disk surface is in the small dust branch).
It seems that the $f_{\mbox{\scriptsize beam}}$ values do not have mirror symmetry with respect to the central position, which may be due to structures in the high-density midplane (e.g., spiral arms or crescent) as reported by \citet{2021ApJ...920...71A} and \citet{2023ApJ...954..190X}.

The higher temperature in the disk mid-plane than in the surface and the $-0.6$--$-0.7$ power-law index in the radial temperature profiles suggest the effect of non-radiative heating mechanisms such as viscous heat dissipation.
Similar suggestions have been made in the observational case studies towards the other Class 0/I YSOs \citep[e.g.,][]{Liu2021,Zapmoni2021MNRAS.508.2583Z,2024arXiv240108722T}, as well as towards TMC-1A \citep{2023ApJ...954..190X}.
If this is the case, given certain surface temperature, the mid-plane temperature is positively correlated with the Planck/Rosseland mean opacities and $\Sigma_{\mbox{\scriptsize dust}}$ (\citealt{Chiang1997ApJ...490..368C}).
The vertical temperature profile of the grown dust branch may not be physical due to the small Planck/Rosseland mean opacities of the grown dust. 
Given the present observational data, it is difficult to robustly distinguish the grown and small dust branches, as well as telling whether $q$ should be $-2.5$ or $-3.5$.
Pinpointing these parameters may require either the observations at more frequency bands or a self-consistent model of dust and gas energetics and heat transportation.

\clearpage

\subsection{Polarization of the two branches} \label{sec:polmodel}
Our SED fitting suggested two branches with small grains ($a_{\rm max}\sim 0.12$~mm) and grown grains ($a_{\rm max}\sim 4$~mm) in the protostellar disk in TMC-1A. Another observational contraint may hint at a difference between the two branches. TMC-1A was observed in linearly polarized continuum emission at the 1.3 mm wavelength \citep{2021ApJ...920...71A}, which is independent of the non-polarized fluxes. To take account of this condition, model images of the polarization Stokes $I$, $Q$, and $U$ are compared with the results of \citet{2021ApJ...920...71A} in this section.

Our model is based on the fiducial model made for TMC-1A by \citet{2023ApJ...954..190X} because our purpose is not to reproduce the observed intensity distribution in detail, exploring various parameters, but to verify whether the observed polarization signal prefers either of the two branches. While the density and temperature distributions of the original model is calculated from a mass accretion rate and a Toomre Q parameter based on the idea of gravito-viscous heating \citep{2022ApJ...934..156X}, we adopt power-law radial profiles, for simplicity, that approximate the dust surface density and temperature distributions: $\Sigma_{\rm dust}(R)= 1600~{\rm g~cm}^{-2}~(R/10~{\rm au})^{-1.96}$ and $T_{\rm dust}= 185~{\rm K}~(R/10~{\rm au})^{-1.27}$, where $R$ is the axial radius. The dust temperature is assumed to be vertically isothermal. The dust density is calculated from the dust surface density and the dust scale height same as the gas scale height determined by the hydrostatic equilibrium. In addition, The dust opacity is calculated with a maximum grain size of $a_{\rm max}$ using the DSHARP opacity model \citep{Birnstiel2018ApJ...869L..45B}. The grain size is spatially uniform. Two maximum grain sizes, 0.12 and 4~mm, are tested to verify the small grain and grown grain branches, respectively. 
Other related parameters follow the fiducial model, such as the star$+$disk mass $M_{\rm tot}$, the disk radius $R_d$, and the power-law index of the grain size distribution $q$.
Then, the Stokes $I$, $Q$, and $U$ intensities in the unit of ${\rm Jy~pixel}^{-1}$ are calculated at 1.3 mm through RADMC-3D \citep{2012ascl.soft02015D} by solving radiative transfer. Finally, the model image is convolved with the same 2D Gaussian beam as that of the observed image in \citet{2021ApJ...920...71A}. While the noise level of the observed image is $25~\uJB$, our model does not include any artificial noise.

\begin{figure}[!htpb]
\gridline{
\fig{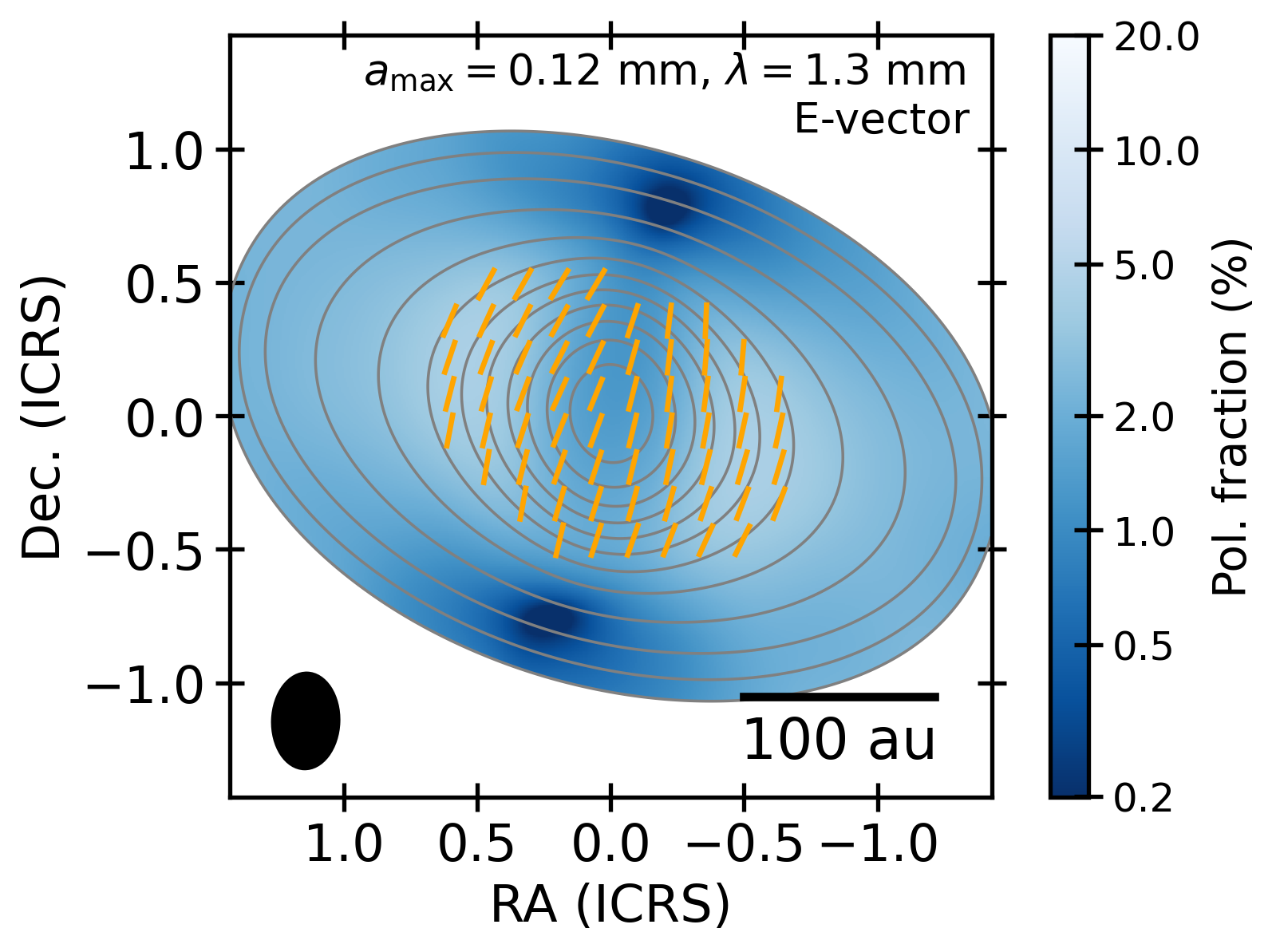}{0.45\textwidth}{(a)}
}
\gridline{
\fig{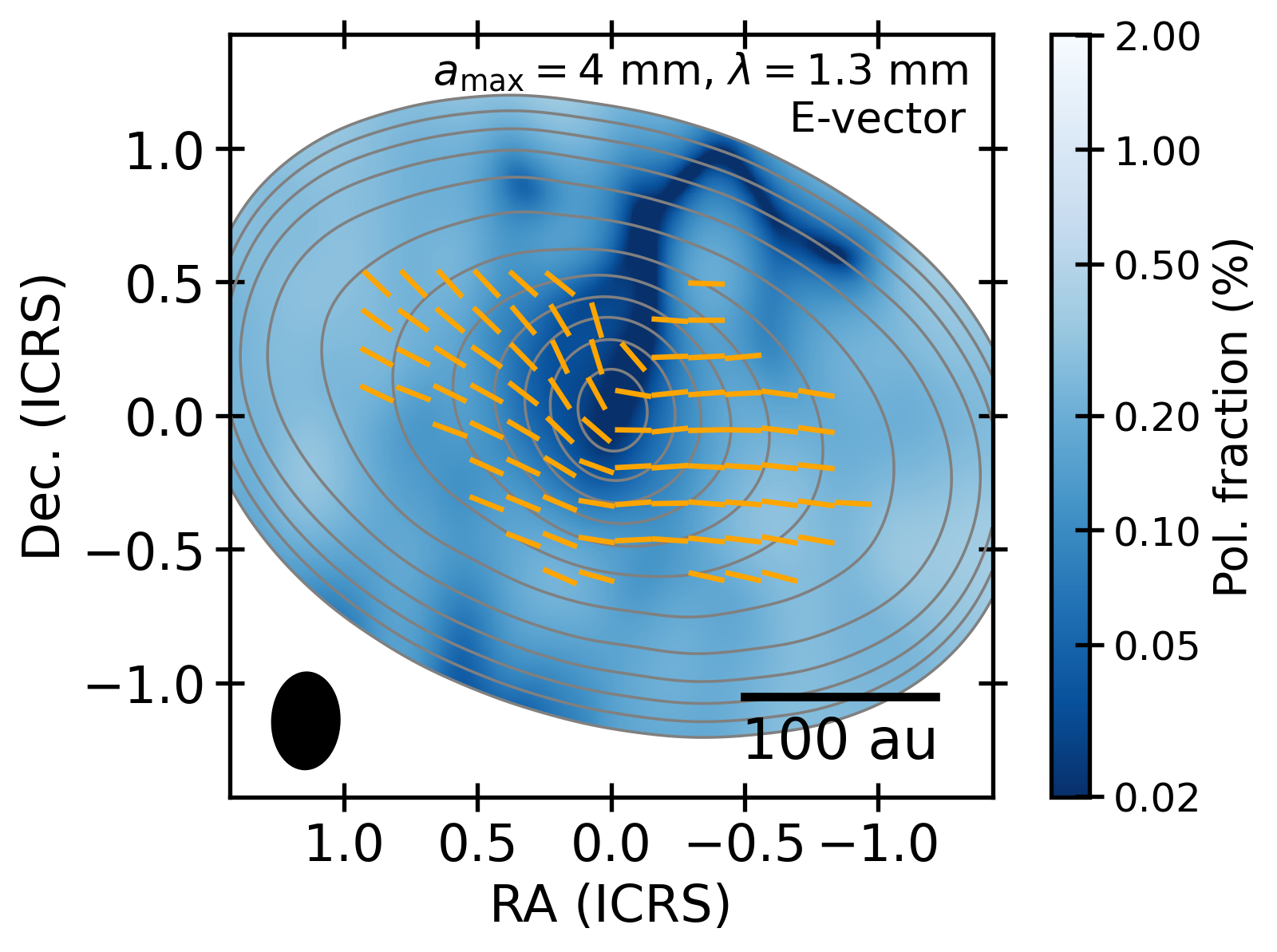}{0.45\textwidth}{(b)}
}
\caption{Disk models with maximum grain sizes of (a) 0.12 and (b) 4~mm. The contours show Stokes $I$, where the contour spacing is $3, 6, 12, 24...\times 25~\uJB$; $25~\uJB$ is the rms noise level in \citet{2021ApJ...920...71A}. The polarziation fraction is shown at positions where Stokes $I$ is $>75~\uJB$. The polarization directions are shown using segments at positions where the polarization intensity is $>75~\uJB$ in panel (a) and $>7.5~\uJB$ in panel (b). The spatial scale, beam size, and contour levels follow Figure 2 of \citet{2021ApJ...920...71A}.}
    \label{fig:modelpol}
\end{figure}

Figure \ref{fig:modelpol}(a) shows results of the small-grain model in the polarized 1.3-mm continuum emission. The spatial scale and the color range follows Figure 2 of \citet{2021ApJ...920...71A}. The model image shows consistent results with the observation: (1) The polarized intensity is detected above $75~\uJB$ in the central $r\lesssim 50$~au, (2) The polarization fraction is $\sim 1\%$ in the central region, (3) The polarization direction is mainly in the disk minor-axis direction inside the central region, and (4) The Stokes $I$ intensity is consistent except for the peak that is roughly twice higher than the observation. The central strong emission of our model may be attributed to the vertically isothermal temperature, the uniform grain size distribution, or the dust model; however, adjusting Stoke $I$ is beyond the scope of this section, i.e., the comparison of polarized signals.

Figure \ref{fig:modelpol}(b) shows results of the grown-grain model. While the spatial scale is the same, the color range is ten times lower than that of Figure \ref{fig:modelpol}(a). The Stokes $I$ intensity of this model is similar to that of the small-grain model. The other points are, however, different from those of the observations and the small-grain model: (1) the polarized intensity is below $75~\uJB$ in the central region, (2) The polarization fraction is $\lesssim 0.05\%$, and (3) The polarization direction is not in the disk minor-axis direction, even if the Stokes $Q$ and $U$ components below the detection level ($75~\uJB$) are considered.

\section{Discussion} \label{sec:discussion}
\subsection{Effect of free-free emission} \label{sec:free}
Fluxes at centimeter wavelengths in protostellar systems could include free-free emission from protostellar jets \citep{1998AJ....116.2953A, 2013ApJ...775...63D, 2014ApJ...780..155L, 2015ApJ...801...91D, 2018ApJ...852...18T, 2019A&A...631A..58C}. The NRAO VLA Archive Survey Images Page\footnote{https://www.vla.nrao.edu/astro/nvas/} shows no detection in a VLA observation toward TMC-1A at the C band (6.1 cm), where free-free emission is expected, with a beam size of $0\farcs 36$ and a sensitivity of $0.07~\mJB$. While the sensitivity is $\sim 10$ times worse than that of \citet{2018ApJ...852...18T} with a similar beam size. The detected C band fluxes of the protostars in \citet{2018ApJ...852...18T} are $\gtrsim 0.6~\mJB$ if they were located at the distance of TMC-1A. Hence, the existing VLA C band data suggest that the free-free emission in TMC-1A is weaker than other protostars. This little impact of free-free emission is supported by two other evaluation methods in this subsection.

The first method uses the bolometric luminosity. Previous studies suggest the correlation between the free-free emission at 4.1 and 6.4 cm and the bolometric luminosity \citep{2018ApJS..238...19T}. Using this relation and power-law scaling, \citet{2022ApJ...934..156X} evaluate the contribution of free-free emission at the VLA Ka band. The same scaling can be adopted for TMC-1A. \citet{2018ApJS..238...19T} show empirical laws of $\log L_\mathrm{4.1cm}=(-2.66\pm 0.06) + (0.91\pm 0.06)\log L_\mathrm{bol}$ and $\log L_\mathrm{6.4cm}=(-2.80\pm 0.07) + (1.00\pm 0.07)\log L_\mathrm{bol}$, where $L_\mathrm{6.4cm}$ and $L_\mathrm{4.1cm}$ are the luminosity (distance-corrected flux) of free-free emission at each wavelength in the unit of mJy~kpc$^{2}$, while $L_\mathrm{bol}$ is the bolometric luminosity in the unit of $\Lsun$. Assuming the same spectral index from 6.4 to 0.9 cm, the free-free luminosity at 0.9 cm can be estimated as $\log L_\mathrm{0.9cm}=(-2.66\pm 0.06) + (0.6\pm 0.3)\log L_\mathrm{bol}$. Then, the bolometric luminosity of TMC-1A, 2.3~$\Lsun$ \citep{2023ApJ...951....8O}, and its distance, 141.8 pc, provide the expected free-free flux of $F_\mathrm{0.9cm}=0.18\pm 0.05~\mJ$. This flux is an order of magnitude smaller than the actual observed flux, supporting that the Ka band flux is dominated by the dust emission in TMC-1A.

The second method is the continuum morphology. The free-free emission tends to be elongated in the outflow direction in a protostellar system \citep{2018ApJ...852...18T}. Our results show that the elongated direction of the Ka band emission in TMC-1A is P.A.$=89\arcdeg \pm 6\arcdeg$ (Section \ref{sec:qk}) based on the 2D Gaussian fitting. This direction is closer to the disk major axis \citep[P.A.$=75\arcdeg$;][]{2021ApJ...920...71A} than the disk minor axis or outflow/jet direction. This morphology supports that the VLA Ka band traces dust emission in TMC-1A.

\subsection{Grain size in other protostellar disks}

The grain size measurements in protostellar disks are limited to a few targets yet (Section \ref{sec:intro}). Previous observational studies suggest several tens $\micron$ for the disk of the Class 0 protostar HH212 \citep{2021ApJ...910...75L}, 100~$\micron$ for the outside and 1~mm for the inside of the soot line ($r=10$~au) in the disk of the Class 0 protostar IRAS 16293-2422 B \citep{2024A&A...682A..56Z}, $>1$~mm for the late Class I protostar WL 17 \citep[][Hashimoto et al. in prep.]{2023ApJ...956....9H}, $>1$~mm for the disk of the Class I/$\II$ protostar HL Tau \citep{2023ApJ...953...96Z}, and $>3$~mm for the outer region in the disk of Class I/$\II$ protostar DG Tau \citep{2023ApJ...954..110O}. Our results correspond to the middle of these Class 0 and I protostars in terms of the evolutionary phase. While a natural interpretation may be that dust grains grow to $\sim 100~\micron$ during the Class 0 phase and grow father to $> 1$~mm during the Class I phase, concluding the time scale of grain growth requires to resolve the grain size distribution in protostellar disks, evalutate the evolutionary stage of each protostellar system more precisely, and increase the sample of protostellar disks where the grain size is estimated through polarization and non-polarization methods using multiple wavelengths and considering the optical depth.

\subsection{Disk mass}
With the multi-layer model, the small grain branch has larger surface densities than the grown grain branch by around an order of magnitude. In other words, the prefernce of the small grain branch means that a higher disk mass is preferred. Previously \citet{Harsono2018NatAs...2..646H} argued that a disk mass high enough to explain an absorption feature in the C$^{18}$O $J=2-1$ line should induce a non-axisymmetric structure in the disk. While the authors declined the possibility of such a high disk mass because the TMC-1A disk does not show a non-axisymmetric structure obvioously, recent observations have identified a one-armed spiral-like structure as a residual from axisymmetric disk models in the ALMA Bands 6 and 7 \citep{2021ApJ...920...71A, 2023ApJ...954..190X}. Furthremore, \citet{2023ApJ...954..190X} has explained the pitch angle of the spiral by gravitational instability of the TMC-1A disk. \citet{2021ApJ...920...71A} also show that the pitch angle is consistent with the ratio between the infall velocity and the rotational velocity along the spiral. These recent results support the possibility of a disk mass high enough to induce gravitational instability. Gravitational instability in the Class I phase could also be the reason of the small grain size, as the turbulence driven by gravitational instability suppresses collisional growth of typical dust grains \citep{2023ApJ...946...94X, 2023ApJ...954..190X}. As a result, planet formation may need to wait until the infall-driven gravitational instability stops, and may occur during the transition between Class I and $\II$ phases. This is consistent with a recent survey result that protostellar disks appear to show less substructures \citep{2023ApJ...951....8O} than more evolved protoplanetary disks \citep{2018ApJ...869L..41A}.

\subsection{Complementarity of analyses}
The comparison of the polarized emission between the observation and our two models (Section \ref{sec:polmodel}) suggests that the small grain branch ($a_{\rm max}\sim 0.12$~mm) is preferable to the grown grain branch ($a_{\rm max}\sim 4$~mm). The two branches show different behaviors in a frequency range between the VLA Q band and ALMA Band 6, which has not been observed yet with the coherent condition. Other frequencies higher than the ALMA Band 7 may also be required to trace the temperature distribution more directly. In the case of self-scattering, the polarization fraction as a function of the observed wavelength can contrain the maximum grain size \citep[e.g.,][]{2017ApJ...844L...5K}. Hence, new observations at different wavelengths but with the coherent condition, in polarized and non-polarized emission, are necessary to further constrain the grain size in this prosotar at the Class I stage, where planet formation is expected to start.

The model-independent SED fitting is complementary to analytic disk models, such as the one built by \citet{2023ApJ...954..190X}.
On one hand, our SED fitting demonstrates that even multi-wavelength dust continuum observations that spatially resolve the disk is insufficient to uniquely pin down dust properties. 
On the other hand, analytic disk models rely on the assumption of a constant gas-to-dust mass ratio throughout a disk. This can be achieved when the dust grains remain small and thus are dynamically well coupled with the gas in the disk.
The consistency between our model-independent SED fitting (the small grain branch) and the analytic disk model of \citet{2023ApJ...954..190X} implies that this assumption is reasonable, in TMC-1A, and validates the physical assumptions of the analytic disk model. After such validation, an analytic disk model has strengths that it can fit the entire disk structures with free parameters as few as those of the model-independent SED fitting.
In contrast, analytic disk models may work only in diffuse regions with small grains when there are azimuthally (a)symmetric local concentrations of grown grains \citep[e.g., ring or crescent; ][]{2024arXiv240202900L} that are largely different from a rather smooth disk assumed in analytic disk models. Even in such a case, the model-independent SED fitting provides complementary diagnostics of dust growth and migration. It is thus important to develop both analyses together with new observations.


\section{Conclusions} \label{sec:conclusion}
We have observed the Class I protostar TMC-1A using the VLA at the Q (7 mm) and Ka (9 mm) bands and analyzed archival data of ALMA at Bands 6 (1.3 mm) and 7 (0.9 mm) at an angular resolution of $\sim 0\farcs 2$ to accurately estimate the grain size in the protostellar disk. The main results are summarized below.
\begin{enumerate}
\item
The VLA images show a compact structure, in both bands, with a size of $\sim 0\farcs 18$ (25 au) and a spectral index of $\sim 3$. The ALMA images show an extended structure with a size of $\sim 2\arcsec$ (300 au), as well as a comapct component. The spectral index calculated from the ALMA images is $\sim 2$ at the central $\sim 40$~au, while it is $\sim 3$ in the outer region.
\item
We searched for models that reproduce the spectral energy distribution derived from the four images at seven different positions. As a result, our SED fitting revealed two branches that are degenerate: a small grain branch with $a_{\rm max}\sim 0.1$~mm and a grown grain branch with $a_{\rm max}\sim 4$~mm.
\item
We also made model images of the polarization fraction and direction and compared the model result with an observed polarization result. The small grain branch provides Stokes I intensity, polarization direction, and polarization fraction more consistent with the observation than the grown grain branch.
\item
The $\sim 0.1$-mm grain size is an intermediate value among the present measurements of the grain size in protostellar systems: Several tens to 100 $\mu$m in systems younger than TMC-1A, while $>1$-3~mm in systems more evolved than TMC-1A. The small grain also implies gravitational instability, suggested by a spiral-like component recent identified. Gravitational instability may suppress planet formation until the transition from the Class I phase to the Class $\II$ phase.
\end{enumerate}


\begin{acknowledgments}
This paper makes use of the following ALMA data: ADS/JAO.ALMA\#2015.1.01415.S, ADS/JAO.ALMA\#2015.1.01549.S, and \\ADS/JAO.ALMA\#2018.1.00701.S. ALMA is a partnership of ESO (representing its member states), NSF (USA) and NINS (Japan), together with NRC (Canada), MOST and ASIAA (Taiwan), and KASI (Republic of Korea), in cooperation with the Republic of Chile. The Joint ALMA Observatory is operated by ESO, AUI/NRAO and NAOJ.
The National Radio Astronomy Observatory is a facility of the National Science Foundation operated under cooperative agreement by Associated Universities, Inc.
H.B.L. is supported by the National Science and Technology Council (NSTC) of Taiwan (Grant Nos. 111-2112-M-110-022-MY3).
\end{acknowledgments}

%

\vspace{5mm}
\facilities{ALMA, VLA}


\software{astropy \citep{2013A&A...558A..33A,2018AJ....156..123A},  
          Numpy \citep{VanDerWalt2011}, 
          CASA \citep[v6.5; ][]{2007ASPC..376..127M},
          emcee \citep{Foreman-Mackey2013PASP},
          corner \citep{corner},
          RADMC-3D \citep{2012ascl.soft02015D},
          }



\appendix

\section{Comments on ALMA data calibration} \label{appendix:almadata}

\subsection{2015.1.01415.S}\label{sub:2015.1.01415.S}
There were three epochs of observations (Table \ref{tab:almadata}).
The pointing and phase referencing centers for the target source TMC-1A were $\alpha({\rm J2000})=$ 04$^{\mbox{\scriptsize{h}}}$39$^{\mbox{\scriptsize{m}}}$35$^{\mbox{\scriptsize{s}}}$.2, $\delta ({\rm J2000})=$ +25$^{\circ}$41$'$44$\farcs$27.
The observations employed five spectral windows that the central frequency and bandwidths (in GHz units) are spw0: 230.538 / 0.117, spw1: 231.469 / 1.875, spw2: 217.055 / 0.469, spw3: 219.561 / 0.117, and spw4: 220.399 / 0.117, respectively.
The ALMA data archive only provided the calibration scripts for the observations taken on 2015 October 23 and 30 since the observations taken on 2015 October 16 failed to pass quality assurance 2 (QA2).
We manually re-calibrated all three epochs of observations using the CASA software package.

The complex gain calibrator adopted in this project (J0440+2728) was relatively faint at the observing frequency ($\sim$50 mJy).
For this reason, all the five spectral windows were combined to maximize the S/N ratios when deriving the complex gain solutions.
The residual gain phase errors appeared large in all three epochs of observations after the basic calibrations.
Both the faintness of the complex gain calibrator and the compromised quality of the gain phase solutions led to the uncertain gain amplitude solutions for the spectral windows spw0, 2, 3, and 4.
We did not find a way that can suppress the amplitude errors without systematically biasing the absolute flux scales.
Hence, only the data taken from spw1, which have low amplitude errors, is used in the spectral index analyses.

We calibrated the absolute flux scales at the central frequency of spw1 by referencing the interpolated flux densities of J0510+1800 measured from the ALMA Calibrator Grid Survey and the SMA Calibrator List.
Specifically, the ALMA Bands 3, 6, and 7 observations and the SMA 225 GHz observations taken on 2015 October 15 were interpolated to calibrate the absolute flux of the observations taken on 2015 October 16; the ALMA Bands 3 and 7 observations taken on 2015 October 23 were interpolated to calibrate the observations taken on the same date; the ALMA Bands 3, 6, and 7 observations and the SMA 225 GHz observations taken on 2015 November 01 were interpolated to calibrate the observations taken on 2015 October 30.
After the absolute flux calibration, the bootstrapped flux densities of the complex gain calibrator J0440+2720 on 2015. Oct. 16, 23, and 30 are ($F_{\mbox{\tiny 217.06 GHz}}=0.075$~Jy, $F_{\mbox{\tiny 231.47 GHz}}=0.056$~Jy), ($F_{\mbox{\tiny 217.06 GHz}}=0.058$~Jy, $F_{\mbox{\tiny 231.47 GHz}}=0.050$~Jy), and ($F_{\mbox{\tiny 217.06 GHz}}=0.058$~Jy, $F_{\mbox{\tiny 231.47 GHz}}=0.051$~Jy), respectively.
Due to the rapid flux variability of J0510+1800, only the absolute flux scale of the observations taken on 2015 October 23 was considered highly reliable.
Nevertheless, given that the bootstrapped flux density of J0440+2728 on 2015 October 30 was closely consistent with that on 2015 October 23, the absolute flux scaling in this epoch of observations is likely accurate, although this does not rule out the possibility that the consistency of the J0440+2728 flux densities on different dates was merely a coincidence.
We do not have a firm mechanism to assess the absolute flux accuracy of the 2015 October 16 observations, although the bootstrapped flux density of J0440+2728 at 231 GHz appeared reasonable. 

To suppress the residual phase errors, self-calibration was performed with several iterations only for gain phase using the CASA \texttt{gaincal} and \texttt{tclean} tasks. 
Limited by the signal-to-noise ratio of the target source, the reasonable minimum solution interval (i.e., without being subject to massive data flaggings) was `per scan'.
The intra-scan phase dispersion of the observations taken on 2015 October 16 was large, which cannot be fully removed by applying the `per scan' gain phase self-calibration solutions.
To avoid biasing the intensity due to phase decoherence, the long-baseline data taken from the observations on 2015 October 16 were not used.
The self-calibration procedures inevitably flagged some long-baseline data.
For all three epochs of observations, we manually flagged some antennae that were located at relatively remote stations to suppress the effects of phase decoherence further.

\subsection{2015.1.01549.S}\label{sub:2015.1.01549.S}
There was an epoch of observations taken on 2016 July 26 (Table \ref{tab:almadata}).
The pointing and phase referencing centers for the target source TMC-1A were $\alpha ({\rm J2000})=$ 04$^{\mbox{\scriptsize{h}}}$39$^{\mbox{\scriptsize{m}}}$35$^{\mbox{\scriptsize{s}}}$.2, $\delta ({\rm J2000})=$ +25$^{\circ}$41$'$44$\farcs$44.
The observations employed five spectral windows that the central frequency and bandwidths (in GHz units) are spw0: 329.986 / 1.875, spw1: 340.724 / 0.059, spw2: 339.352 / 0.059, spw3: 329.340 / 0.059, and spw4: 330.598 / 0.059, respectively.
The absolute flux, passband, and complex gain calibrators are the quasar J0510+1800.
We manually re-calibrated the observations using the CASA software package.
When deriving the complex gain solutions, all the five spectral windows were combined to maximize the S/N ratios.

We calibrated the absolute flux scales by referencing the ALMA Calibrator Grid Survey measurements on J0510+1800.
In the ALMA Calibrator Grid Survey, 91.5/103.5 GHz measurements were carried out two days before the observations on TMC-1A, and 343.5 GHz measurements were carried out two days after the observations on TMC-1A.
These measurements were interpolated to the central frequency of spw0.
Due to the rapid flux variability of J0510+1800, we expect the uncertainty of the absolute flux after calibration to be $\sim$30\%.
We do not think this epoch of ALMA observations can yield reliable intra-band spectral indices and thus only used the data taken at spw0 in our scientific analyses.
There were no complementary SMA observations on J0510+1800 on proximate dates, which prevents us precisely assessing the absolute flux and spectral index errors.

The residual gain phase errors were large after the basic calibrations.
To suppress the phase errors, self-calibration was pwerformed with several iterations only for gain phase using the CASA-\texttt{gaincal} and \texttt{tclean} tasks. 
Only the broadband continuum window spw0 was used with the per-integration solution interval of 6.05 seconds.

\subsection{2018.1.00701.S}\label{sub:2018.1.00701.S}
This was a polarization track carried out on 2018 November 21. 
Details of the data calibration can be found from \citet{2021ApJ...920...71A}.
The absolute flux calibrator J0510+1800 was covered by the ALMA Calibrator Grid Survey on 2018 November 21 (at multiple frequencies).
Therefore, the absolute flux scale of this track is very reliable.

\begin{deluxetable*}{l c c c c c c}
\tablecaption{Summary of the ALMA archival data\label{tab:almadata}}
\tablehead{
\colhead{Date} &
\colhead{Band} &
\colhead{PWV\tablenotemark{a}} &  
\colhead{{\it uv}-range} &
\colhead{Mean freq. (bandwidth)} &
\colhead{Flux calibrator\tablenotemark{b}} &
\colhead{Self-cal. solint\tablenotemark{c}}
\\
\colhead{(UTC)} &
\colhead{} &
\colhead{(mm)} & 
\colhead{(meter)} &
\colhead{(GHz / GHz)} &
\colhead{(Jy, ---)} &
\colhead{(second)} 
} 
\startdata
\multicolumn{7}{c}{Project Code: 2015.1.01415.S (c.f. \citealt{Bjerkeli2016Natur.540..406B,Harsono2018NatAs...2..646H})} \\
2015 Oct. 16 &
6&
2 & 
64--16105 &
231.5 / 1.875 &
$F_{\mbox{\tiny 223.81GHz}}=2.53$, $\alpha=-0.26$ &
$\sim50$\\
2015 Oct. 23 &
6&
0.6 & 
80--14720 &
231.5 / 1.875 &
$F_{\mbox{\tiny 223.81GHz}}=2.71$, $\alpha=-0.28$ &
$\sim50$\\
2015 Oct. 30 &
6&
0.4 & 
79--15186 &
231.5 / 1.875 &
$F_{\mbox{\tiny 223.80GHz}}=3.22$, $\alpha=-0.24$ &
$\sim50$\\
\hline
\multicolumn{7}{c}{Project Code: 2015.1.01549.S (c.f. \citealt{Harsono2021AA...646A..72H})} \\ 
2016 Jul. 26 &
7&
0.4 & 
14--1051 &
330.0 / 1.875 &
$F_{\mbox{\tiny 334.00GHz}}=1.32$, $\alpha=-0.47$ &
6.05\\
\hline
\multicolumn{7}{c}{Project Code: 2018.1.00701.S (c.f. \citealt{2021ApJ...920...71A})} \\
2018 Nov. 21 &
6&
0.8 & 
14--1256 &
233.0 / 2.000 &
$F_{\mbox{\tiny 224.11GHz}}=1.68$, $\alpha=-0.50$ &
6.05\\
\enddata
 \tablenotetext{a}{Mean precipitable water vapour.}
 \tablenotetext{b}{The flux density of the absolute flux calibrator J0510+1800, at a certain frequency, that we assumed during the data reprocessing, shown with the spectral index $\alpha$ of the calibrator at the frequency. It is derived based on referencing to either the ALMA calibrator grid survey or the SMA Calibrator List.}
 \tablenotetext{c}{Solution interval for gain phase self-calibration. The intervals of $\sim 50$ and 6.05 second correspond to `per scan' and `per integration', respectively.}
\end{deluxetable*}



\bibliography{sample631}{}
\bibliographystyle{aasjournal}



\end{document}